\documentclass[useAMS,usenatbib]{mn2e}
\pdfoutput=1
\usepackage{amssymb}
\usepackage{graphicx}
\usepackage{float}

\usepackage{color,soul}
\usepackage[inline]{enumitem}
\usepackage{booktabs}

\newcommand{\be}{\begin{equation}}
\newcommand{\ee}{\end{equation}}
\newcommand{\mad}{{MAD}\,}

\newcommand{\msun}{${\rm M}_{\odot}$\,}

\newcommand{\msuny}{${\rm M}_{\odot}\,{\rm yr}^{-1}$\,}

\newcommand{\pjet}{$\rm{P}_{jet}$\,}

\newcommand{\racc}{${\rm r}_{acc}$\,}

\newcommand{\E[1]}{\times10^{#1}}

\newcommand{\mnras}{MNRAS\,\,}
 
\newcommand{\aap}{A\& A\,\,}
\newcommand{\apjl}{ApJL\,\,}
\newcommand{\apj}{ApJ\,\,}

\newcommand{\apss}{Astrophys. Space Sci.\,\,}

\newcommand{\apjs}{Astroph. J. Supp.\,\,}
\newcommand{\nar}{New Astron. Review\,\,}
\newcommand{\pasj}{Pub. Ast. Soc. Japan\,\,}

\newcommand\corr[1]{\normalfont \color{black}#1}
\newcommand\corrx[1]{\normalfont \color{black}#1}

\begin{document}

\title[Backflows in cocoons]{Backflows by AGN jets: Global properties and influence on SMBH accretion}

\author[Cielo et al.]{ S.~Cielo$^{1}$\thanks{e-mail: cielo@iap.fr%
}, V.~Antonuccio-Delogu$^{2}$\thanks{e-mail: Vincenzo.Antonuccio@oact.inaf.it}, J. Silk$^{1,3,4,5}$, and A.D. Romeo$^{6}$ \\
$^{1}$ Institut d'Astrophysique de Paris (UMR 7095: CNRS  UPMC ??? Sorbonne Universit\'{e}s), 98 bis bd Arago, F-75014 Paris, France\\
$^{2}$ INAF/Istituto Nazionale di Astrofisica-Catania Astrophysical Observatory, Via S. Sofia 78, I-95126 Catania, Italy\\
$^{3}$ Laboratoire AIM-Paris-Saclay, CEA/DSM/IRFU, CNRS, Univ. Paris VII, F-91191 Gif-sur-Yvette, France\\
$^{4}$ Department of Physics and Astronomy, The Johns Hopkins University Homewood Campus, Baltimore, MD 21218, USA\\
$^{5}$ BIPAC, Department of Physics, University of Oxford, Keble Road, Oxford OX1 3RH, United Kingdom\\
$^{6}$ Purple Mountain Observatory, 2 Bejing Xilu, 210008 Nanjing, China
}

\date{Accepted ??. Received ??; in original form 20xx ??}

\maketitle
\pagerange{\pageref{firstpage}--\pageref{lastpage}} \pubyear{2016}

\label{firstpage} 
\begin{abstract}

Jets from Active Galactic Nuclei (AGN) inflate large cavities in the hot gas environment around galaxies and galaxy clusters.
The large-scale gas circulation promoted within such cavities by the jet itself gives rise to backflows that propagate back to the centre of the jet-cocoon system, spanning all the physical scales relevant for the AGN. 

Using an Adaptive Mesh Refinement code, we study these backflows through a series of numerical experiments, aiming at understanding how their global properties depend on jet parameters. We are able to characterize their mass flux down to a scale of a few kiloparsecs to about $0.5$~\msuny for as long as $15$ or $20$~Myr, depending on jet power. We find that backflows are both spatially coherent and temporally intermittent, independently of jet power in the range  $10^{43-45}$~erg/s.

Using the mass flux thus measured, we model analytically the effect of backflows on the central accretion region, where a Magnetically Arrested Disc lies at the centre of a thin circumnuclear disc. Backflow accretion onto the disc modifies its density profile, producing a flat core and tail.

We use this analytic model to predict how accretion beyond the BH magnetopause is modified, and thus how the jet power is temporally modulated. Under the assumption that the magnetic flux stays frozen in the accreting matter, and that the jets are always launched via the Blandford-Znajek (1977) mechanism, we find that backflows are capable of boosting the jet power up to tenfold during relatively short time  episodes (a few Myr).

\end{abstract}

\section{Introduction: backflow morphology and AGN jet self-regulation}

The propagation of AGN jets inflates large, hot, turbulent cavities in the interstellar medium of their host galaxies. Circulation of gas in such cavities gives rise to pronounced streams of hot gas flowing back from the hot spot (if present, as in FRII radio galaxies), along the cavity boundaries to the central plane. 

Such \emph{backflows} are driven by the thermodynamics of the gas, and --- once in the central plane --- consist of very low angular momentum gas, which potentially reaches down to very small scales, contributing to  the mass and energy supply in  the accretion region around the SMBH. Backflows carry very hot, high pressure gas; they can thus heavily affect \emph{circumnuclear star formation} and the properties of the accretion disc, as a self-regulating feedback mechanism.
Backflows as a feature of jet-cocoon systems were already noticed in the first simulations of the propagation of relativistic jets into homogeneous atmospheres \citep{1982A&A...113..285N}, and confirmed in more recent simulations \citep{2008A&A...488..795R,2007MNRAS.382..526P}. \cite{2010ApJ...709L..83M} distinguish between two types of backflows, according to the different geometries of the flow itself: a \emph{straight} backflow, with flow lines extending from the tip of the hotspot back to the origin, and a \emph{bent} backflow, where the flow lines acquire curvature.
In early 2D simulations, precursors of what we present in this work, \cite{2010MNRAS.405.1303A} also noticed the formation of this feature, and that the backflow evolved from a \emph{bent} to a \emph{straight} geometry. In {\corr that} work, backflow was described as large-scale vorticity created by sharp gradients in the thermodynamic state of the gas at the hot spot and cavity boundaries, precisely as stated by a fundamental theorem of fluid dynamics, known as \emph{Crocco's theorem} \citep{1937ZaMM...17....1C}. This can be understood from the Euler momentum equation:
\be 
\frac{\partial\mathbf{v}}{\partial t} -  \mathbf{v}\times\nabla\times\mathbf{v} = -\nabla h +T\nabla S \label{eq:crocco:1}
\ee
Here S is the entropy and $h = U + p/\rho +v^{2}/2\,$ is the \emph{stagnation enthalpy} \citep{2001...2100..25stzmatost, 1992pavi.book.....S}. Even for a stationary flow, Crocco's theorem states that vorticity can only be created by finite gradients of enthalpy $h$ and/or entropy $S$.
\begin{table*}
 \setlength{\tabcolsep}{5pt} 
\begin{center}
	\smallskip
	\begin{tabular}{lccccccccccc}
	\hline\hline
\multicolumn{3}{c|}{Simulation}      & \multicolumn{2}{c|}{Halo}	& \multicolumn{5}{c|}{Jet} & \multicolumn{2}{c|}{Backflowing mass} 	\\
\multicolumn{3}{c|}{ }               & \multicolumn{2}{c|}{    }	& \multicolumn{5}{c|}{   } & \multicolumn{2}{c|}{(at given time)} 	\\
Name & Resolution & $t_\mathrm{max}$ & $M_{200}$   & $t_{cool, 0}$	& $P_\mathrm{jet}$ & $\mathcal{M}_\mathrm{jet}$	& $\Delta t_\mathrm{jet}$ & $\dot{m}$ & $\dot{p}$         & $ Total$    & $central$   \\
	 & [pc]       & [Myr]		     & [$M_\odot$] & [yr]			& [erg/s]		   & 							& [Myr]                   & [\msun/yr]  & [\msun/yr $km/s$] & [$M_\odot$] & [$M_\odot$] \\
\hline		
\multicolumn{7}{l}{\it{Elongated Cavity series}} \\
\hline	
EC42	 & 78.125 & 473	 & $1.7\E[12]$ & $6\E[8]$ & $10^{42}$ & 5 & 79 & 0.0088 & 167.22 & $4.84\E[5]$ & $1.28\E[4]$ \\
		 &        &      &             &          &           &   &    &        &        & 20 Myr      & 20 Myr	   \\          
EC43	 & 78.125 & 140	 & $1.7\E[12]$ & $6\E[8]$ & $10^{43}$ & 5 & 42 & 0.0190 & 776.15 & $4.69\E[5]$ & $7.11\E[3]$ \\
		 &        &      &             &          &           &   &    &        &        & 10 Myr      & 10 Myr	   \\          
EC44	 & 78.125 & 115	 & $1.7\E[12]$ & $6\E[8]$ & $10^{44}$ & 5 & 21 & 0.0409 & 3603   & $9.92\E[5]$ & $1.9\E[4] $ \\
		 &        &      &             &          &           &   &    &        &        & {\corr 7 Myr}      & 7 Myr	   \\          
\hline		
\multicolumn{7}{l}{\it{Round Cavity series}} \\
\hline	
RC44	& 78.125 & 23.1 & $2.6\E[12]$ & $4\E[8]$ & $1.12\E[44]$ & 5	& 5.37 & 0.0237 & 2900  & $6.90\E[4]$ & $1.04\E[5]$ \\
        &        &      &             &          &              &   &        &        &       & 10 Myr      & 10 Myr	  \\    
RC45	& 78.125 & 22.2	& $2.6\E[12]$ & $4\E[8]$ & $1.12\E[45]$ & 5	& 5.37& 0.0510 & 13461 & $4.84\E[4]$ & $6.80\E[4]$ \\
        &        &      &             &          &              &   &        &        &       & 8 Myr       & 8 Myr       \\        
RC46	& 78.125 & 22.2 & $2.6\E[12]$ & $4\E[8]$ & $1.12\E[46]$ & 5	& 5.37& 0.1098 & 62548 & $2.71\E[5]$ & $4.34\E[4]$ \\
        &        &      &             &          &              &   &        &        &       & 7 Myr       & 7 Myr	      \\       \hline 
\end{tabular}
\end{center}
\caption{Parameters, timings and bubble characteristics. All simulation parameters: run specifications (name, smallest cell side, simulation time), halo parameters (mass, central cooling time), jet parameters (power of each of the two jets, Mach number, lifetime, mass and momentum injection fluxes) and total backflow  gas mass at the given time (i.e. the total mass in the backflow region isolated in Figures \ref{fig:large-scaleEC} to \ref{fig:RC}).}
\label{tab:runs}
\end{table*}

\cite{2010MNRAS.405.1303A} pointed to the connection between backflows with large-scale vorticity in the cavity. The flow begins near the hot spot (HS), where \corr{a curved shock front induces a jump in entropy and a gradient in the Bernoulli constant transverse to the shock. The downstream gas thus gains a vorticity \citep{1992pavi.book.....S}}, and \corr{its flow is} then confined between the dense and hot bow shock from the outer side, and the hot turbulent cavity gas from the inner side.

This goes on until the gas falls back to the central plane and follows the cavity edge (or collides with mirror backflows \corr{in a bipolar jet}) falling down towards the jet origin with very low impact parameter (and thus angular momentum), although its inflow velocities reach up to several hundreds km s$^{-1}$. 

In three dimensions, however, this mechanism loses some effectiveness, as with the additional degree of freedom the velocity could be directed (in absence of other constraints)  anywhere in the $z=0$ plane. Also, the flow is subject to more effective hydrodynamic instability, which can slow and disrupt it.

In previous 3D simulations, \cite{Cielo2014} showed that despite the unarguably reduced efficiency, substantial backflows (always around $1$~\msuny) reach the central few hundred parsecs. The duration of such backflows varies with jet power (higher powers move cavities away from the centre at earlier times, killing backflows), but always encompasses a few Myr. Furthermore, the backflow gas was found to be stable against hydrodynamics instabilities, although the simulations covered just the first few million years.

Observations of backflows have been quite challenging for a long time, as the gas is hot but very sparse, and only mildly relativistic, so easily out shined by the jets. However, observational characterization of backflows has recently been emerging; in particular \cite{laing2012backflows} observed backflows in two low-luminosity jetted radio-galaxies. In particular, a mildly backflowing component around the lobes is needed in order to fit the emission and polarization radio maps.


AGN are multi-scale systems, and as the backflows get to closer to the central BH, they experience all the relevant physics at different scales (\citealp{Antonuccio-Delogu:2007oq}): after the thermodynamics-dominated circulation in the cavities, they will collide with a central structure (likely, a circumnuclear-nuclear disc extended up to about $100$~pc), where they may generate further inflows.

After this stage, this secondary inflow may enter the \emph{magnetosphere} of the BH, where the dominating energy source is the magnetic field, ultimately responsible of launching the jets.

Throughout this work, we aim to follow backflows from start to end. This is important when we consider their effect as a source of hot gas accretion onto the central BH, capable of triggering further jet activity or increasing the power of an already running jet in a self-regulating context.

We start from the largest scales (kpc or larger) by making use of the hydrodynamic simulations,  described in Section \ref{sec:setup} and interpret the results on the basis of \emph{Crocco's theorem}. For this purpose, we will provide visual snapshots of the density field, as well as histograms of the gas spatial distribution
along the z-axis (Section \ref{sec:aroundCavities}).

Next, we proceed by investigating, with similar methods, the flow of gas which has already reached the $z = 0$ plane. In order to quantify the impact parameter of the gas at this stage, we plot the evolution of its \emph{circularization radius} and calculate the mass flux within $2$~kpc (Section \ref{sec:centralDisk}).

{\corr We then explore the effects of this infall onto the circumnuclear region, not resolved in our simulations,  assuming that the innermost magnetized structure is a} \emph{Magnetically Arrested Disc} (Section \ref{sec:MADdisk}).

Finally, in Section \ref{sec:self-regulation} we develop some quantitative considerations of how the processes described above influence the rate of mass accretion onto the central SMBH and the production of the jet itself (via the Blandford-Znajek mechanism from \citealp{1977MNRAS.179..433B}). 

\section{Setup and simulation description}\label{sec:setup}

We run our simulation using the hydrodynamic, \emph{Adaptive Mesh Refinement} (AMR) code FLASH v. 4.2 \citep{Fryxell:2000fk}. In our computational setup, we solve the non-relativistic Euler equations for an ideal gas with specific heat ratio $\gamma = 5/3$ (see Appendix \ref{appendix} for more details).
In order to properly model the energetic of the system, we include a static, spherical gravitational potential of the host dark matter halo (following a NFW profile), as well as the self-gravity of the gas. 
This set-up is an evolution  of that used in \citet{Cielo2014} {\corr ---} henceforth C14.

We choose a cubic 3D computational domain and Cartesian coordinates.
The side of the box is in all runs fixed to $L=640$~kpc, much larger than the maximum extent of the jets,
in order to encompass most of the halo, whose typical size $R_{200}$ is $\sim250$~kpc.
The resolution, defined as the size of the smallest computational element, is in all cases fixed to $\Delta r=78.125\, {\rm pc}$ (see Appendix \ref{appendix}). All the boundary conditions are set to the FLASH \emph{outflow} (i.e. \emph{zero-gradient}) default value.
We present two families of runs, mainly differing in jet power $P_{jet}$. In both series, the dark matter halo density follows a spherical NFW profile with $R_\mathrm{200} = 0.25$~Mpc, $M_\mathrm{200} = 1.71\times10^{12}\,M_\odot$ and a concentration parameter $c_\mathrm{200} = 7.8$ \citep{Dutton2014}. We then fill this potential well with hot coronal gas, having a uniform temperature $T_0$ and metallicity ([Fe/H] = -1.0) in hydrostatic equilibrium within the dark matter potential. 
We achieve the latter condition by adopting the following constant temperature profile, 
introduced by \cite{Capelo2013}:
\begin{equation}
 \rho_{g}(r) = \rho_0\,\exp\left(-\mu m_{p}\frac{\left[\Phi(r)-\Phi(0)\right]}{k_B\,T_0}\right)
\end{equation}
where $\mu$ is the mean molecular weight of the gas, $\Phi(r)$ the gravitational potential 
and $k_B$ the Boltzmann constant. For the normalization, we follow \citet{McCarthy2008} in setting the ratio of gas to dark matter mass within the halo $R_\mathrm{500}$ to 0.85 times the cosmic baryon fraction (taken in turn from \citealt{WMAP7}). Once the normalization of the gas profile is fixed, the gas properties are thus completely specified by the chosen $T_0$, or alternatively by the gas central \emph{cooling time}, which we report in Table \ref{tab:runs} (column 5).

The first family is the EC series (for \emph{Elongated Cavities}), where the DM halo has a mass fixed to $1.7\E[12]$\msun and a central cooling time of $6\E[8]$~yrs. For this series, we launch three runs, differing only in the jet mechanical power $P_{jet}$, which assumes values of $10^{42}$, $10^{43}$ and $10^{44}$~erg/s (run EC42, EC43 and EC44, respectively).
The second series (denoted the RC series, for \emph{Round Cavities}), differs from the EC series mostly in that it features a higher halo mass, a slightly shorter cooling time (a consequence of a denser hot gas phase) and higher values of $P_{jet}$ (runs RC44, RC45 and RC46; see Table~\ref{tab:runs} for a complete list of all parameters). 

{\corr The dependence of the cavity's shape on the jet/halo physical parameters was previously investigated in C14. Briefly, the cavity shape is linked to these parameters via the jet's thermal pressure: a jet having higher thermal pressure (or propagating in a colder environment) will originate rounder cavities as the over-pressure determines isothermal expansion; on the contrary a \emph{cold} jet with a small ratio between internal and kinetic pressure will inflate elongated  cavities. This being said, in our case the RC jets create rounder cavities than their EC counterparts for three reasons:
\begin{itemize}
\item they are intrinsically hotter, as they are more powerful (i.e. faster) but at injection they have all the same \emph{internal Mach number} $\mathcal{M}_{jet}=5$ (in other words, the Mach number relative to the environment changes).
\item the core of the halo in which they propagate is slightly colder;
\item they thermalize a larger part of their total energy, as more powerful jets have shorter lifetimes, and thermalization is more efficient at early stages.
\end{itemize}
}

\begin{figure*}
\begin{centering}
\includegraphics[width=0.95\textwidth]{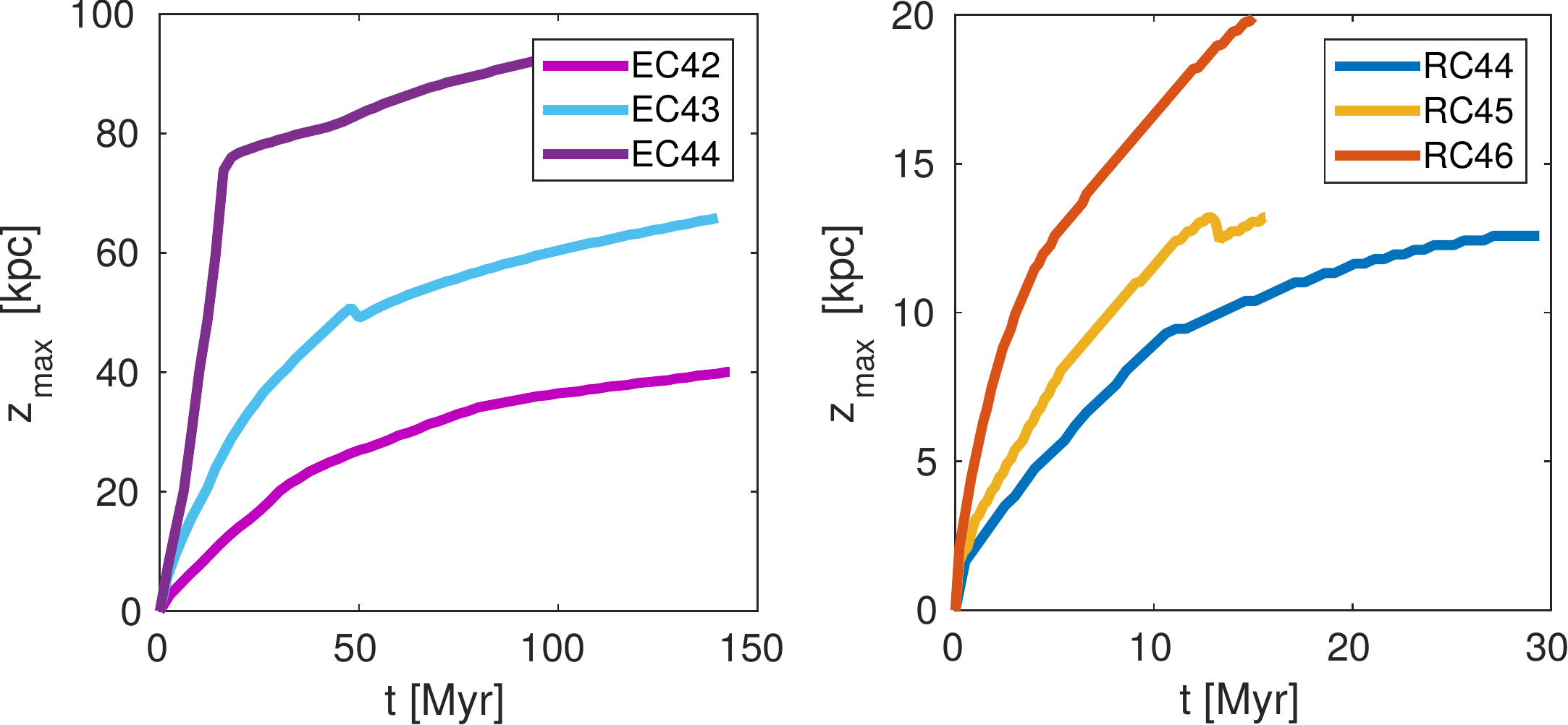}
\caption{Location of the bow-shock (i.e., jet advancement) with time. LEFT: EC (for Elongated Cavities) series. RIGHT: RC (for Round Cavities) series. Velocities are of order $1000$~km/s in most cases, although the highest power jets of each series can reach up to tenfold higher velocities, e.g. EC44 before the jet switches off).} \label{fig:advance}
\end{centering}
\end{figure*}

The jets are modelled by injecting gas in the central region for a prescribed lifetime.  The injection power, mass and momentum flux are also reported in Table \ref{tab:runs} for each run (columns 6, 9 and 10, respectively); a given kinetic power also corresponds to  a specific  lifetime $\Delta t_{jet}$ (column 8 of Table \ref{tab:runs}). In the EC series, backflows on the central plane significantly fade before \corr{the end of the jet's active phase}; in the RC series, however, there are residual backflows even after the jets are switched off. \corr{In some cases the injection velocities resulting from these parameters choices are slightly superluminal. The ram pressure of the local environment where the jet emerges, however, brings the jet to non-relativistic velocities in the first few cells\footnote{We {\corr had also tried increasing} the jet's cross section and lowered the velocity to keep the injected power constant, but the results are indistinguishable after sufficiently long times.}}. {\corr The overall jet/Hot Spot advancement velocity is however much slower than that (maximum about $1.5\times{10}^4$~km/s, Figure \ref{fig:advance}), so this high nominal injection speed does not affect the evolution of the cavities.}

The jets advance through the Interstellar Medium (ISM), initially producing a hot, localized shock in the \emph{Hot Spot} (HS) and a bow-shock region, accordingly to the previous findings of C14. In Figure~\ref{fig:advance} we show the location $z_{max}$ of the bow-shock region, simply defined as the geometrical extent along the jet axis of the hot spot;
this gives an idea of the actual advance speed of the jets and of their effect on the hot gas. 

The jets expand up to a maximum z-distance ranging from a few tens up to about $100$~kpc in the EC runs (with a clear turnover when the jets are switched off), while they reach about $20$~kpc in the RC ones, mainly due to the shorter simulation time. In any case, in this work ,we focus on the first few tens of Myr, as backflows arise at these times.

\section{Galactic-scale backflows around cavities} \label{sec:aroundCavities}

\begin{figure*}
\begin{centering}
\begin{minipage}[l]{1\columnwidth}
\includegraphics[width=0.93\columnwidth]{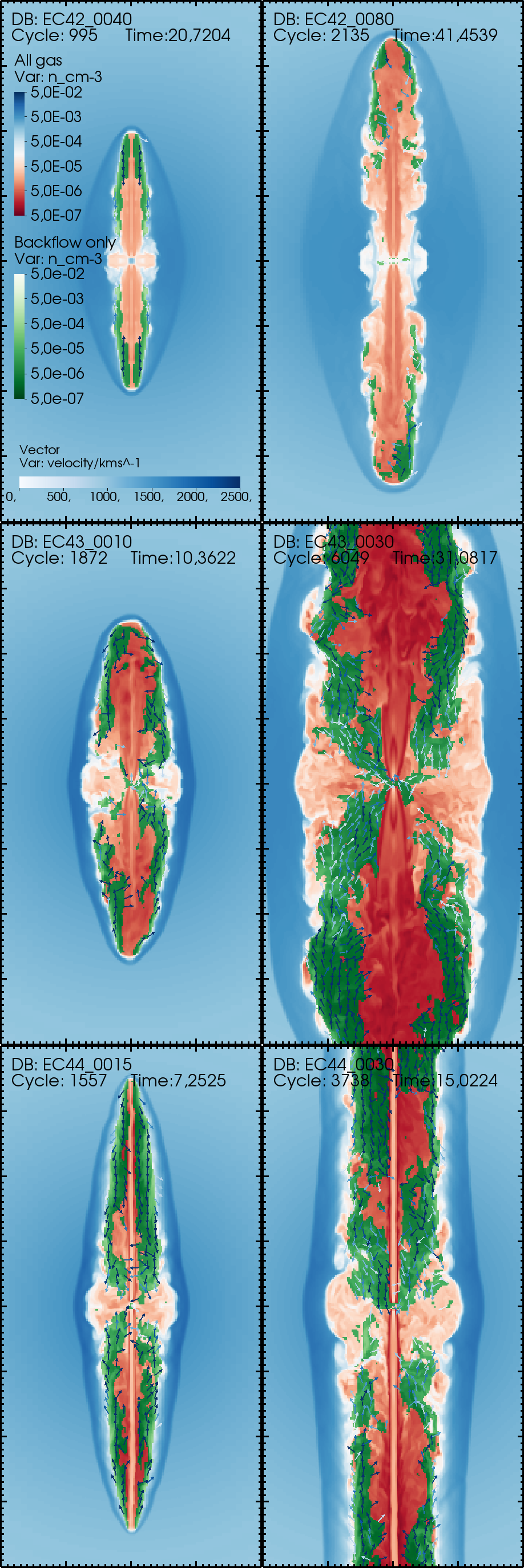}
\end{minipage}
\begin{minipage}[r]{\columnwidth}
\includegraphics[width=0.93\textwidth]{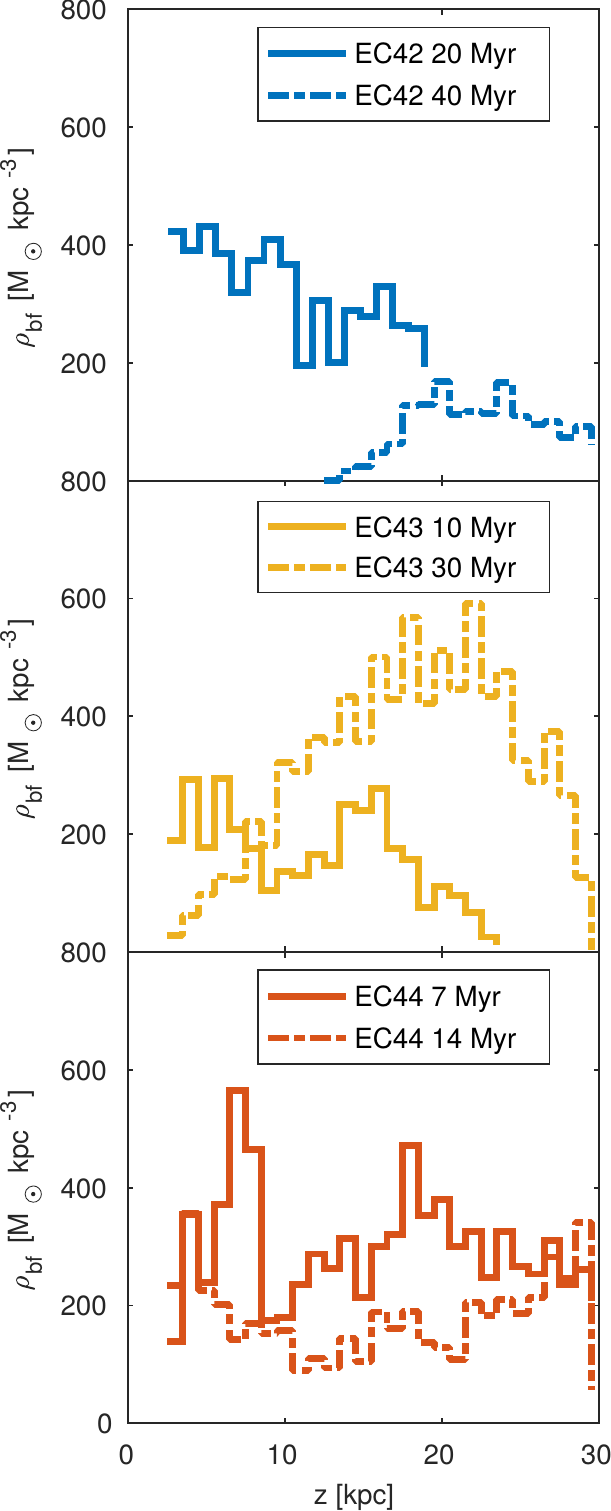}
\caption{LEFT: 40x80~kpc number density slices (in $cm^{-3}$) along the $y=0$ plane for the EC runs (increasing power from top to bottom, increasing time from left to right). The backflow regions  ($\bmath{v}\cdot\bmath{\hat{r}}\leq 0$) are highlighted in green (colour palette on the top-right of each plot), and have the velocity field superimposed (velocity is given in $km/s$ and follows the bottom colour legend of each panel).
RIGHT: density histograms (with a fixed bin amplitude of $500$~pc) of the z-distribution of the same backflowing gas; i.e. the total mass over the total volume of the part of the green region that falls in each bin (in $M_\odot\,{kpc}^{-3}$).
Run name and time are indicated in each panel. Note how backflows move farther out the central region with increasing time.} \label{fig:large-scaleEC}
\end{minipage}
\end{centering}
\label{fig2}
\end{figure*}

\begin{figure*}
\begin{centering}
\begin{minipage}[l]{1\columnwidth}
\includegraphics[width=0.93\columnwidth]{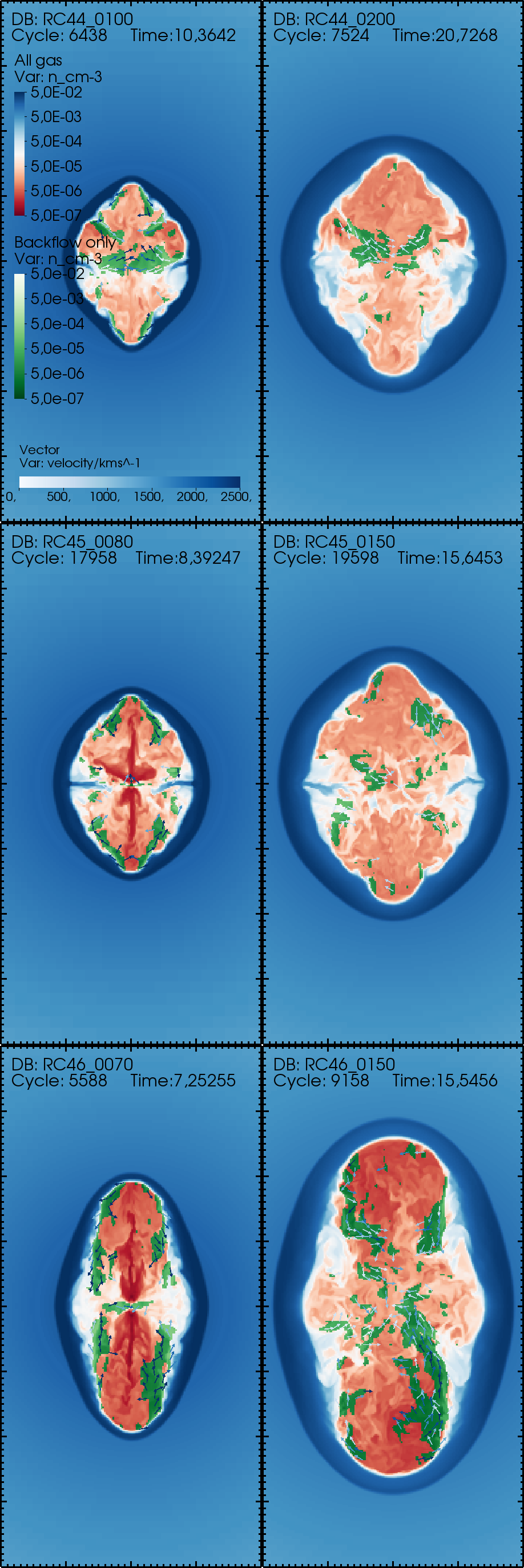}
\end{minipage}
\begin{minipage}[r]{\columnwidth}
\includegraphics[width=0.90\textwidth]{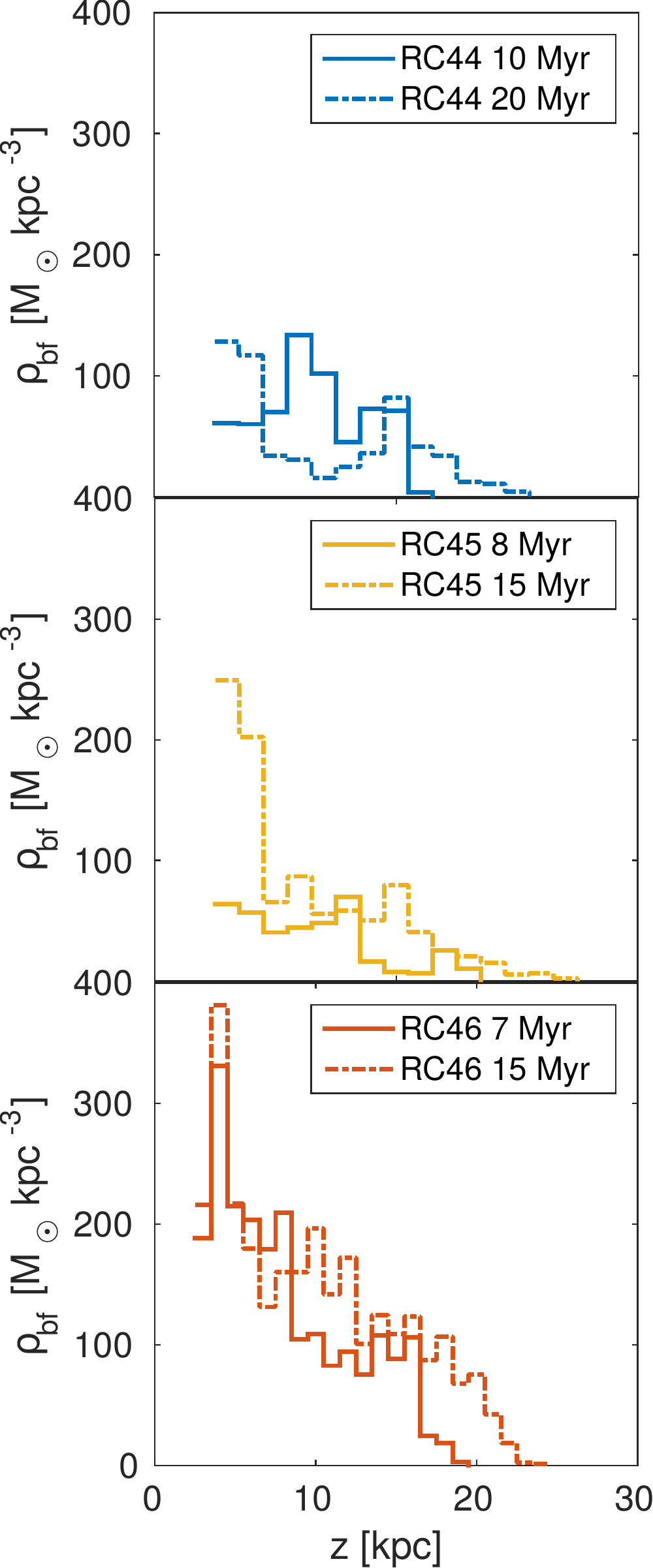}
\caption{Same as Figure \ref{fig:large-scaleEC} for the RC runs. The contours show number density central (y=0) slices of the total and backflowing gas, and corresponding density histograms along the z axis. The RC series has rounder cavities, but also higher jet power and shorter lifetimes. Total mass and duration of backflows in the lobes are reduced compared to EC series (note the reduced scale on the y axis).
The path of the backflows changes too, since it follows more closely the cavity's shape.} \label{fig:large-scaleRC}
\end{minipage}
\end{centering}
\label{fig3}
\end{figure*}

The prediction of Crocco's theorem that backflows originate from the HS and then flow along the lobe/bubble boundary is verified in the simulations run, as can be seen by looking at Figures \ref{fig:large-scaleEC} and \ref{fig:large-scaleRC}\footnote{Some animations of the simulations are available from https://blackerc.wordpress.com/people/salvatore-cielo/}, where we show visual slices of the density field along the $y=0$ plane (left columns). The backflows follow a different colour scheme (colour legend on the right), in order to highlight them within their context in the cocoon; the backflow region is also recognizable as it is the only one where the velocity field is superimposed\footnote{See the colour legend at the bottom of each panel in the Figure for the magnitudes of the arrows.}. 

We define the backflow regions to include all grid cells whose velocities point towards the jet origin within a $\pm45$ degrees cone. 

This (rather conservative) selection is necessary to view a cleaner flow, as backflows are otherwise contaminated by gas patches that `bounce" on the cavity walls. While this is contemplated in our Crocco theorem description, the resulting velocity field shows many spurious features. The constraints we can put on the mass flows with this analysis are for this reason only lower limits.\\
We set an additional threshold on the radial (w.r.t.  the origin) velocity: $v_r\leq-225$~km/s. This is to exclude coherent cooling flows (or occasional highly-turbulent spots) at late times; this \corr{selection} does not significantly change the backflow mass at the {\corr epochs considered in this study}. 

Cooling flows however can develop after about $50$ or $100$ Myr, i.e. later than the time intervals during which the jets are on. A discussion of the cooling flows is outside the scope of this paper; as we see from the slices shown, backflows reach high velocities (often $\leqslant-2500$~km/s) so this selection 
does not exclude any significant part of the backflows even at late times.

Finally, we cut out the innermost $\pm2$~kpc of gas from the cavity selection (Figure \ref{fig:large-scaleEC} and \ref{fig:large-scaleRC}), which will be the object of Section \ref{sec:centralDisk} (as one can see from Figure \ref{fig:diskEC} and \ref{fig:diskRC}).

The backflows initially appear as thin layers contained within the cavity/dense-shell interface (left panels in figures~\ref{fig:large-scaleEC} and~\ref{fig:large-scaleRC}); thus in 3D these flow layers wrap the entire inner cavity. Since the cavities at this stage reproduce the lobes observed in radio-galaxies (see C14), backflows can appear around most $10-20$~kpc galactic radio-lobes.
After $10-20$~Myr, the cocoon develops an internal structure: the lobes detach from the central plane, and leave a gap 
filled by denser gas.
{\corr Following C14, we call this \emph{the lobe phase}. During this phase,}
turbulence develops as a  consequence of shocks and shearing between the different gas layers, which creates turbulent eddies through Kelvin-Helmholtz (KH) instabilities.
 
The large-scale backflows are affected by this structure,  
{\corr and can converge back to the jet axis following the bubble boundaries} (right panels in Figure \ref{fig:large-scaleEC} and \ref{fig:large-scaleRC}). In their path, they {\corr also} take part in the cocoon's turbulent motion, both near the HS and along the shearing cavity boundary; they also contribute to {\corr generate the shear, as they initially consist of laminar flows} in relative motion with respect to both the inner cavities and the outer bow-shock. 

These aspects were 
analysed by C14 and the flows were found to be stable against KH instability\footnote{In C14 the resolution was slightly better than in the present work, however the simulations lasted only $\sim6$~Myr or less.}.
{\corr Regardless of how much they contribute }
to the generation of turbulent motions, the backflows are perturbed and fragmented by 
{\corr it} 
: one can clearly see patches of coherent inward radial velocity in Figure \ref{fig:large-scaleRC},  more prominent at later times (panels on the right) and within the innermost $\sim5$~kpc, where the cavities start to detach from the centre. 

The backflows can gain vorticity and momentum at two different sites: first at the HS, where they start their journey around the lobes, and later near the 
{\corr $z=0$}
plane, since after the 
{\corr lobe phase}
backflows 
{\corr must bend again, this time 
following the jet beam \emph{chimney} 
that connects the lobe to the jet origin. }

Near the central region,
{\corr the backflow } follows instead a rather straight pattern. 
{The mass of the gas 
\corr involved in the backflows (Table \ref{tab:runs}, column 12)}, as well as the time {\corr it takes to get back to the central plane}
depend on the cavity shape and size, and on the hotspot pressure (enthalpy) which gives the initial kick. 

After \corr{a sufficiently long time}, the HS are usually too far away \corr{from the centre, and} the backflows stop halfway. 
{\corr A few Myr after the jet has been switched off ($t \geq \Delta t_{jet}$)
, the lobes turn into roughly spherical bubbles and detach completely from the centre. 
We refer to this stage as the \emph{bubble phase}. 
At this time, cooling flows 
can occur near the $z=0$ plane.  Any residual backflows will then cease; however 
{\corr analogous circulation patterns} will persist in the bubble for all its lifetime (bubbles from light, supersonic jets such as these create vortex-ring-shaped cavities; see e.g. \citealp{Guo2016Light}).
}

In the EC runs, large-scale backflows are more
{\corr extended and carry more mass};
although the jet power is on average lesser, it can drive the gas around the cavities efficiently (also because jet lifetimes are correspondingly longer). 

In the RC runs instead, the presence of more spherical cavities \corr{forces} streamlines to gain more curvature since the start, until they reach the $z=0$ plane.
Also the flow appears more fragmented in the RC case, as more disconnected patches are clearly visible in the slices. This is probably due to the increased turbulence in the cocoon environment generated by hotter and more powerful jets. Such patches linger for up to about $10$~Myr near or within the central plane.\\
\noindent
In order to estimate the backflow location and mass transport, in Figure \ref{fig:large-scaleEC} and \ref{fig:large-scaleRC}, we add density histograms of the backflowing gas distribution along the z-axis, for the same snapshots shown in the slices; the total mass in the region for the first snapshot is also reported in Table \ref{tab:runs}.
In many EC runs, one sees that initially the distribution extends back to the first central few kpc and then moves  farther away at later times. This is less true for the RC runs, although overall the mass involved is smaller by a factor $5$ or $10$.\\
As for the total gas mass \corr{accumulated at the centre}, it seems to be approximately constant (about $5\times10^5\ \mathrm{M}_\odot$ in the EC case, $5\times10^4\ \mathrm{M}_\odot$ in the RC case), except for the most powerful (and shortest-lived) jet events, in which it increases by a large factor (about $2$ and $5$, respectively). 

\section{Kiloparsec-scale backflows on the central disc}\label{sec:centralDisk}

\begin{figure*}
\begin{centering}
\begin{minipage}[l]{1\columnwidth}
\includegraphics[width=0.95\columnwidth]{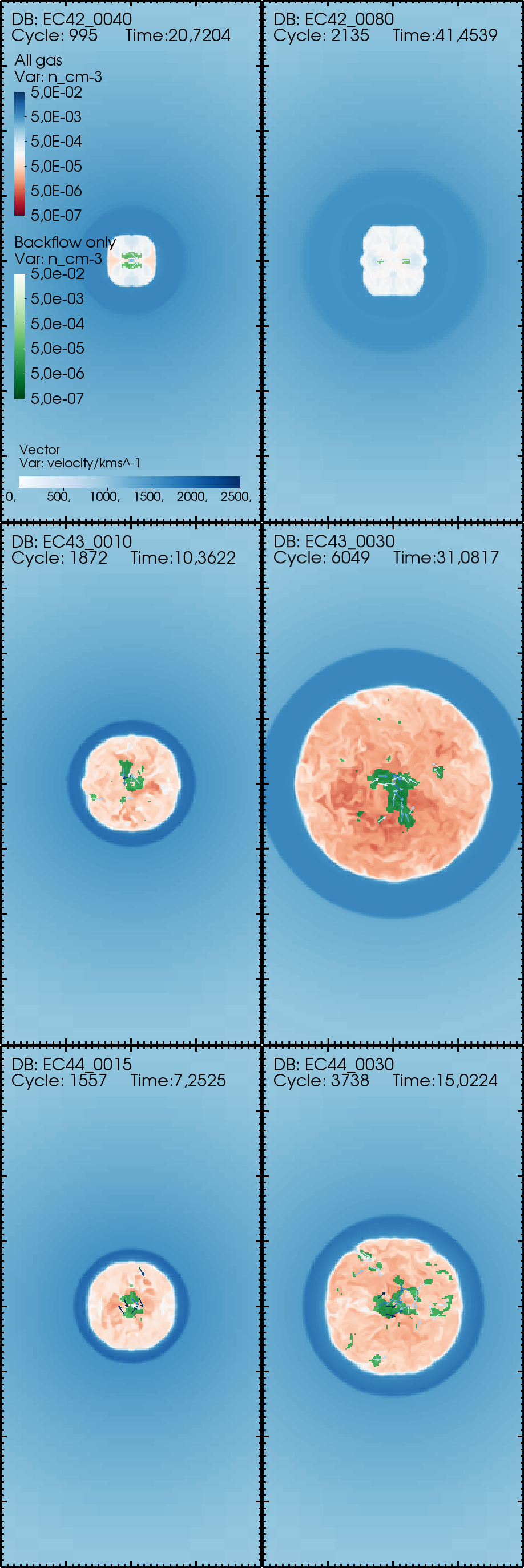}
\end{minipage}
\begin{minipage}[r]{\columnwidth}
\begin{flushright}
\includegraphics[width=0.9\textwidth]{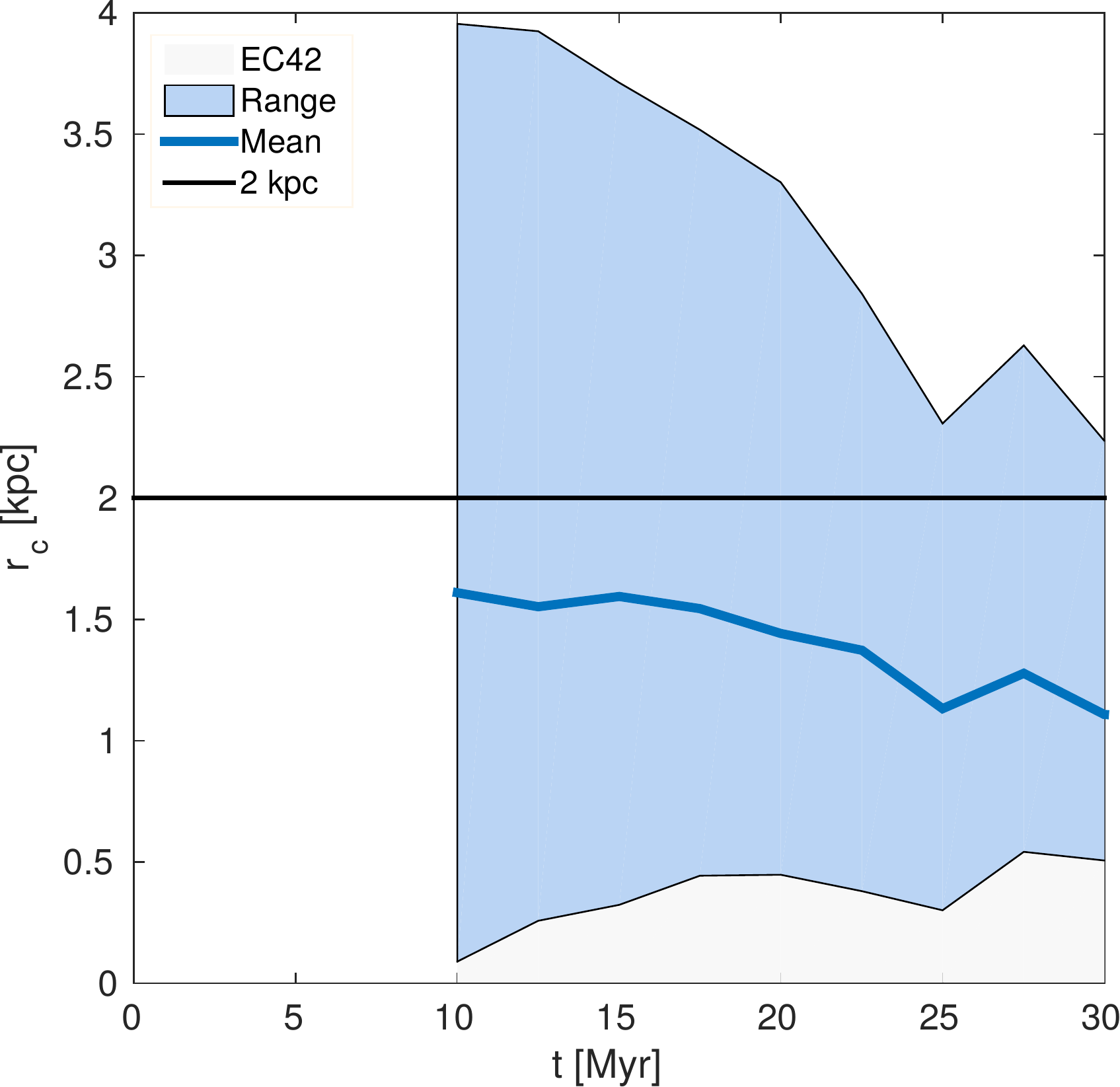}
\includegraphics[width=0.85\textwidth]{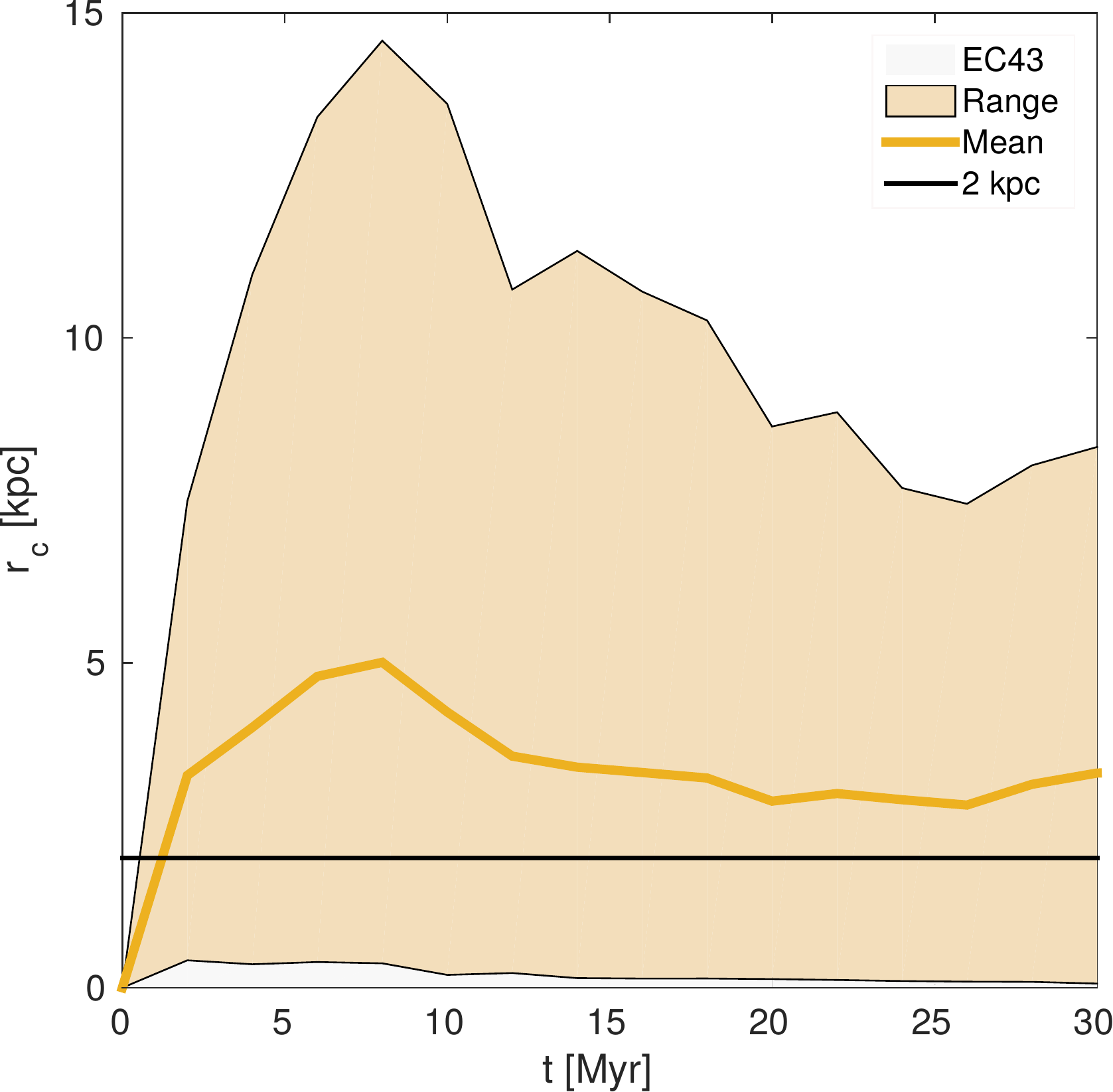}
\includegraphics[width=0.85\textwidth]{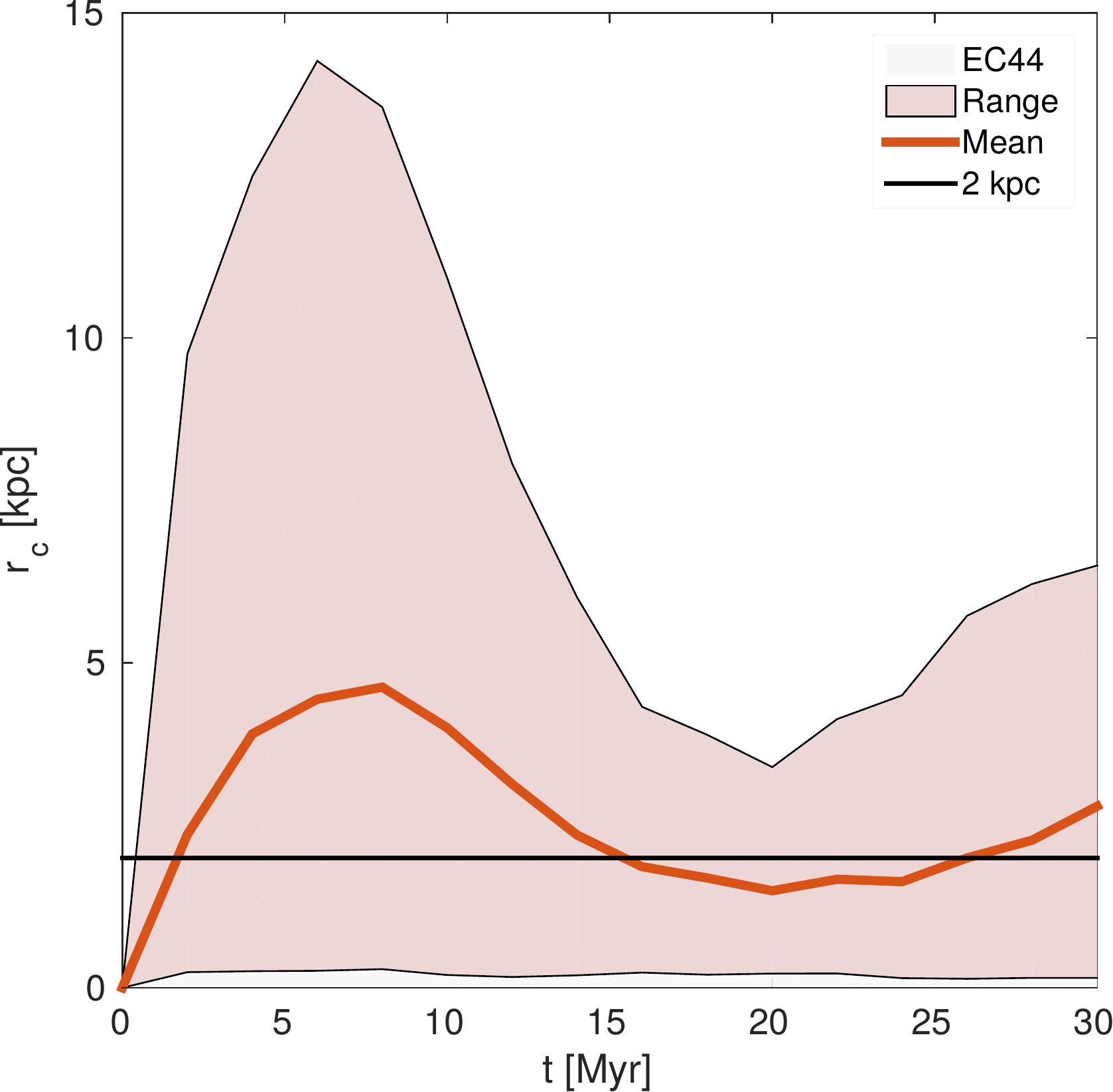}
\end{flushright}
\caption{LEFT: Gas density slices along the $z=0$ plane; similar to Figure \ref{fig:large-scaleEC}, but for the gas in the central region.  RIGHT: time evolution of \emph{circularization radius} $r_c$ of the gas within the control cylinder of radius $2$~kpc and height $0.2$~kpc as mass-average (thick lines) and min-max range (shaded area). Large backflow volumes often show $r_c<2$~kpc (black line). Values smoothed for clarity.
}
\label{fig:diskEC}
\end{minipage}
\end{centering}
\end{figure*}

\begin{figure*}
\begin{centering}
\begin{minipage}[l]{1\columnwidth}
\includegraphics[width=0.945\columnwidth]{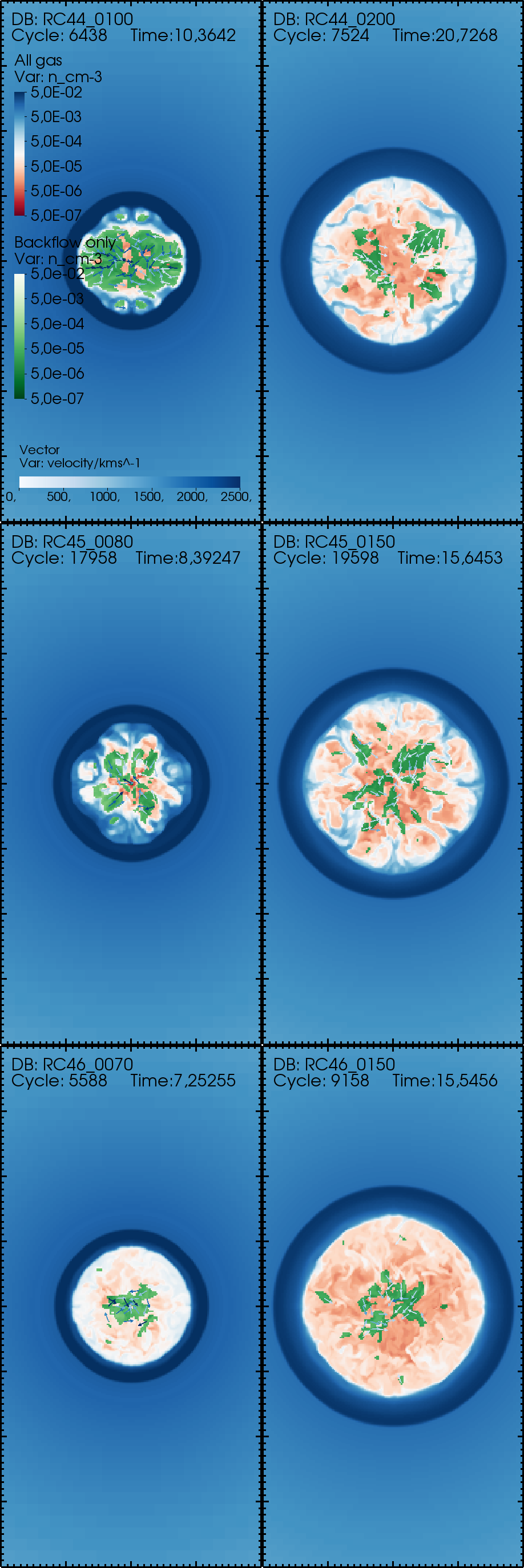}
\end{minipage}
\begin{minipage}[r]{\columnwidth}
\begin{flushright}
\includegraphics[width=0.85\textwidth]{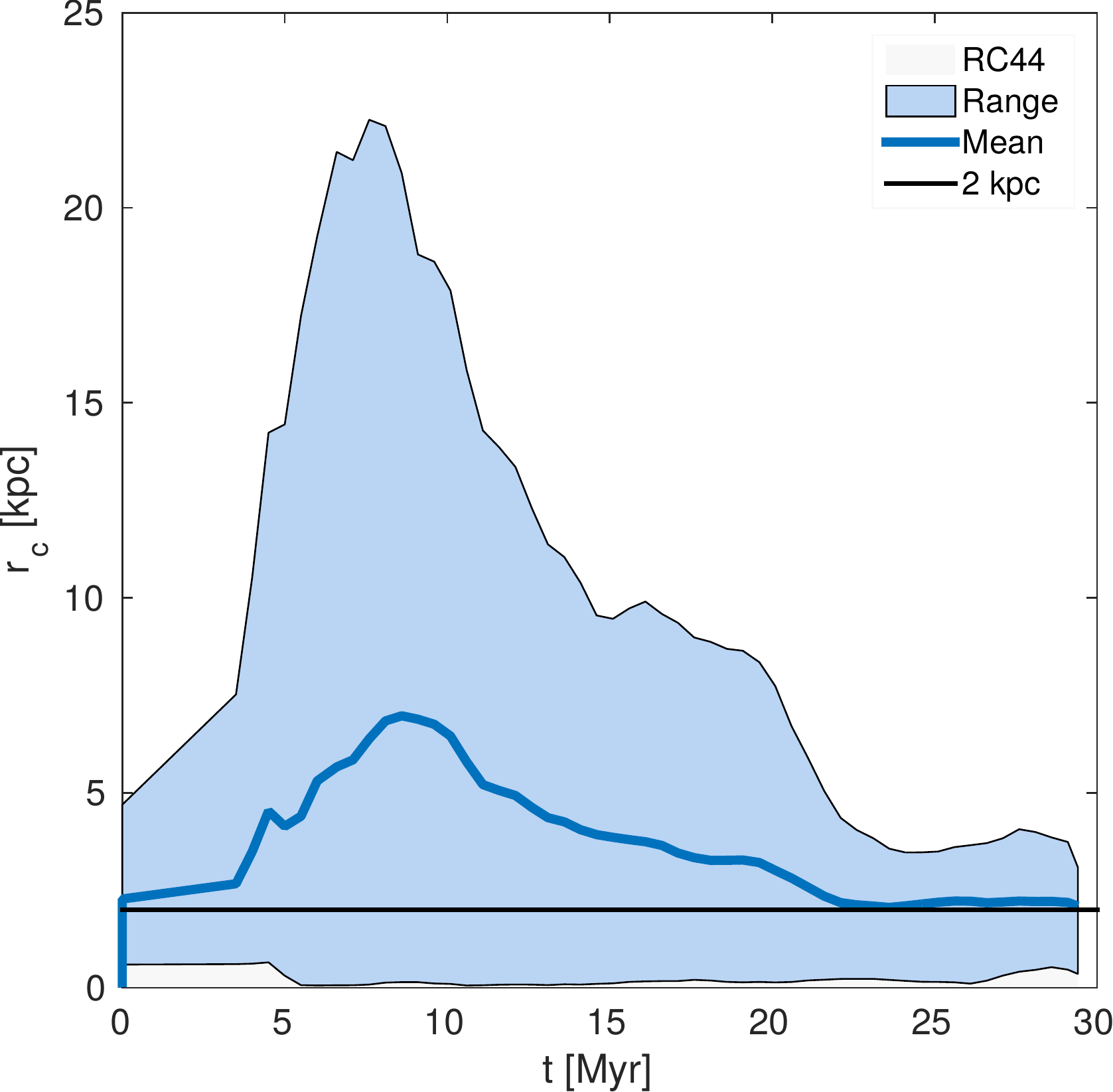}
\includegraphics[width=0.85\textwidth]{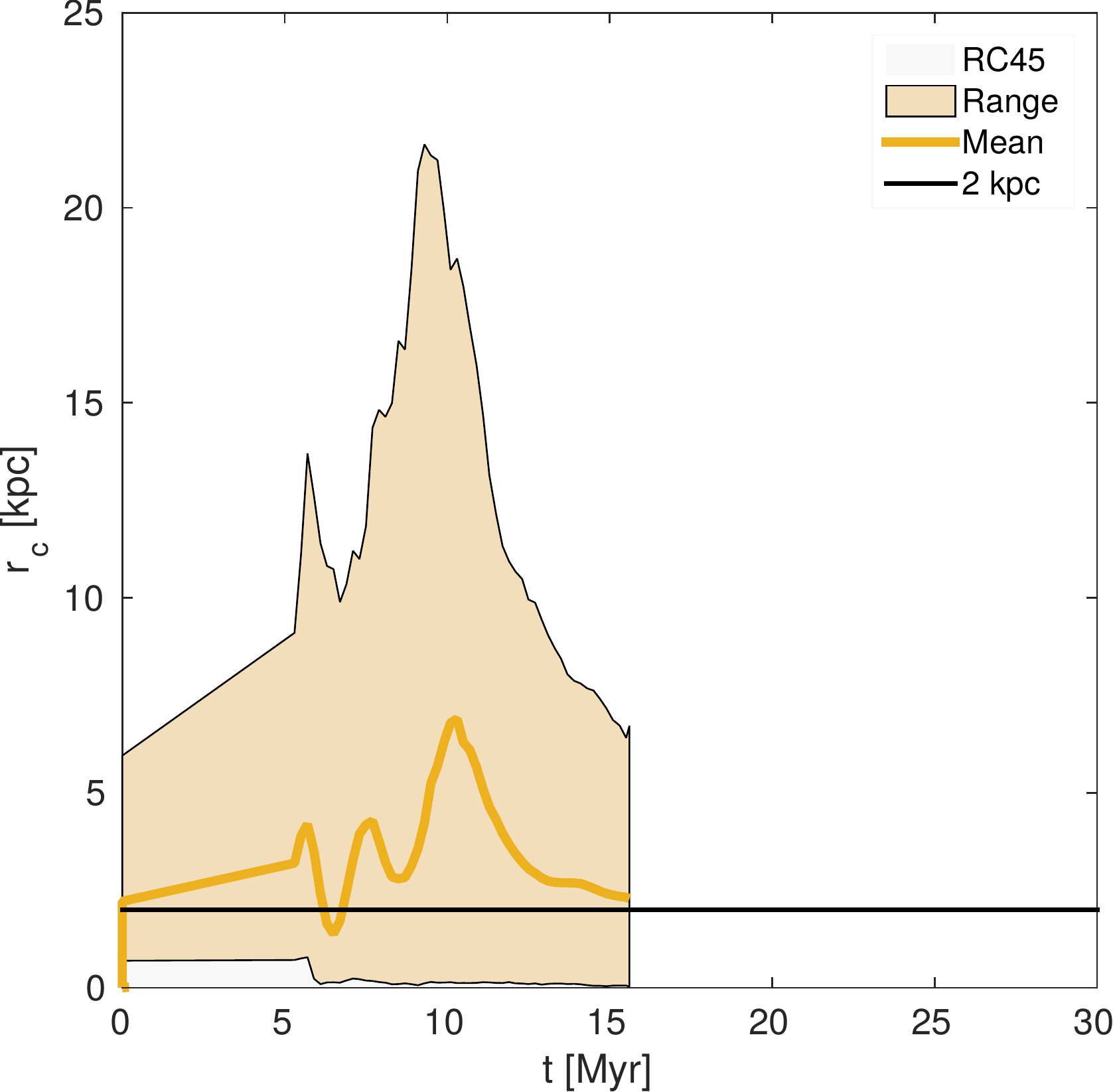}
\includegraphics[width=0.85\textwidth]{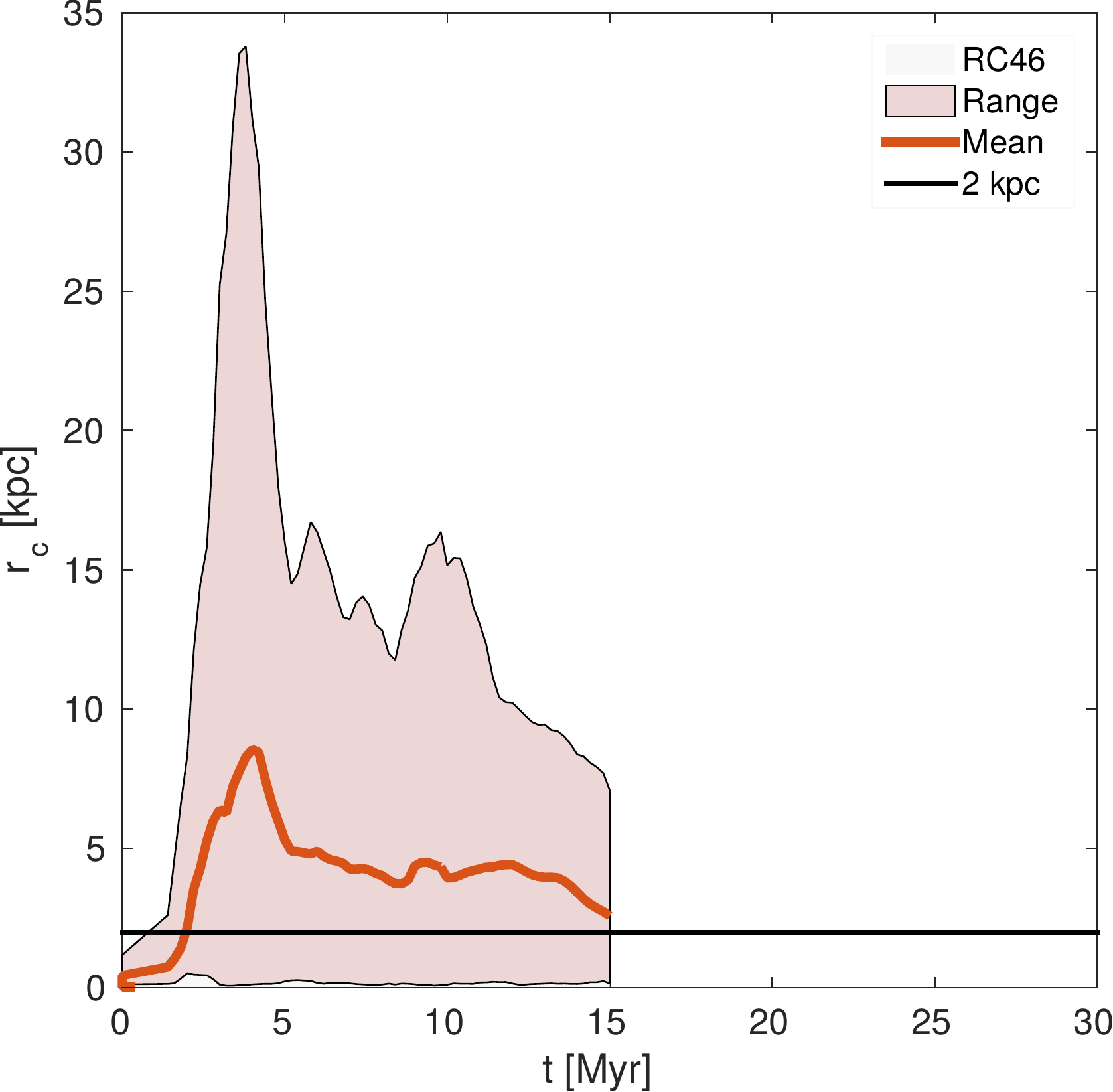}
\end{flushright}\caption{Same as figure \ref{fig:diskEC} for  the RC run family (notice also the different vertical scale); qualitatively, the results are similar, as there is always some gas with low $r_c$, although most often the mass-averaged $r_c$ is larger than $2$~kpc. } \label{fig:diskRC}
\end{minipage}
\end{centering}
\end{figure*}

We now turn our attention to the central region. In Figures \ref{fig:diskEC} and \ref{fig:diskRC} ,we plot density slices along the $z=0$ plane, again with backflowing gas highlighted, and with velocity arrows superimposed. 

The backflow in the central plane is more regular during the first few Myr, and more patchy afterwards, participating in  the turbulence of the cocoon gas that affects the entire cavity. 
{\corr
In this case, the mass transport is significant even for the low power jets. There is more mass involved in central disc backflows in the RC runs than in the EC series (see Table \ref{tab:runs}, last column)} notably different from what is seen for the cavity-wide backflows described in Section \ref{sec:aroundCavities}.

Due to axial symmetry, we expect that the backflowing gas in the central disc should have \corr{on average} little to zero angular momentum, and thus flow directly towards the BH accretion disc.

In order to test this prediction, we need to trace the angular momentum of the gas.
In Figures \ref{fig:diskEC} and \ref{fig:diskRC} (right panels), we plot the evolution of the \emph{circularization radius} --- a proxy for angular momentum --- within a small central cylindrical selection. 

Let $L$ be the modulus of the \corr{specific} angular momentum vector of a gas parcel in a computational cell. We define the circularization radius $r_c$ as the radius the parcel would have if it were on a circular orbit within its host Dark Matter halo:
\begin{equation}
r_{c} = \frac{L^{2}}{GM\left(r_{c}\right)}
\label{eq_rc}
\end{equation} 
Here $M(r)$ is the mass of dark matter\footnote{We neglect self-gravity of the gas, although it is accounted for in the simulations} within $r$. 
The evaluation of equation \ref{eq_rc} is particularly simple as our dark matter halo is spherically symmetric, thus it is straightforward to find the implicit solution of equation~\ref{eq_rc}.
In using the full modulus of the 3D angular momentum of the gas in each cell $L$, rather than its z-component only ($L_{z}\leq L$), we are making a conservative estimate.

We evaluate $r_c$ for all the gas cells of the backflowing gas only, selected with the same velocity threshold as in  Section \ref{sec:aroundCavities}. This time though we select a cylinder centred on the jet origin and having $2$~kpc radius in the xy plane, and a thickness of about $300$~pc (four simulation cells in total) along the z axis. The mass-weighted average value of $r_c$ is plotted for each snapshot time (thick lines in the plots of Figures \ref{fig:diskEC} and \ref{fig:diskRC}) together with the minimum and maximum values in the selected  region (shaded area around the lines) and a $2$~kpc line (in black).
{\corrx In general, gas having $r_c < r$ will always tend to migrate to smaller distances, thus falling towards the central BH. In this case, we can conclude that all gas parcels having $r_c < 2$~kpc will always stay within the selection.}

In {\corr almost} all runs, the average of $r_{c}$ stays above the $2$~kpc line: we interpret this \corr{circumstance as arising from} the fact that 
backflows have  quite high characteristic velocities that  do not necessarily have a negligible impact parameter, thus resulting in {\corr some floor} values for $L$ and $r_{c}$. 
A noteworthy exception is run EC42, in which {\corr backflows reach the central region after $10$~Myr}, while later the average $r_c$ stays always well below $2$~kpc. 
{\corr On the contrary, in} both EC43 and EC44 they grow smoothly up to a relatively early peak at $5$~kpc around $7$~Myr then decline back to $2$ or $3$~kpc. 

\corr{From the figures} we can draw up \corr{two} general conclusions:
\begin{itemize}
\item There is always backflow with $r_c<2$~kpc, as the shaded area always extends down to almost zero.{\corr
\item Statistically, there is always a significant mass fraction able to migrate to smaller radii at all times.}
\end{itemize}
{\corr About the second point, although flow masses are not indicated in the figures, we see from Table  \ref{tab:runs} that the total values are around $10^4$~$\mathrm{M_\odot}$ or a few times that.}
\corr{Note also that angular momentum is not necessarily conserved in the small-scale flows. It could be dissipated through viscosity, or again through the thermodynamic action described by Crocco's theorem, as {\corr patches or streams of gas} from the the opposite sides of the bipolar jet collide in the {\corr $z=0$} plane, as in the 2D-simulations by \citealp{2010MNRAS.405.1303A}; in support of this, the average $r_c$ clearly decreases with time. Thus the values in Figures \ref{fig:diskEC} and \ref{fig:diskRC} are just upper limits on RC.}

\begin{figure}
\begin{flushright}
\includegraphics[width=0.97\columnwidth]{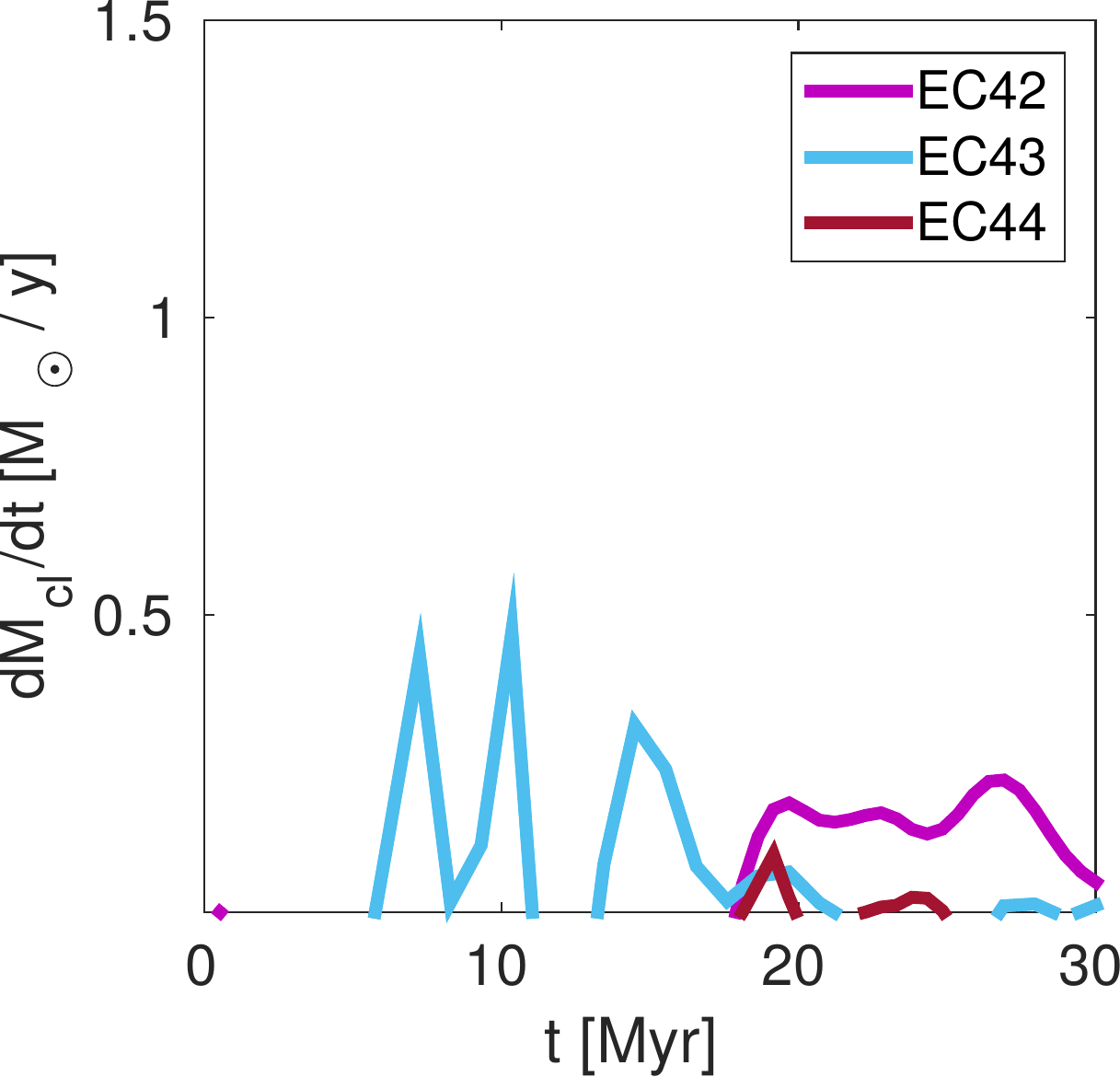}
\includegraphics[width=0.95\columnwidth]{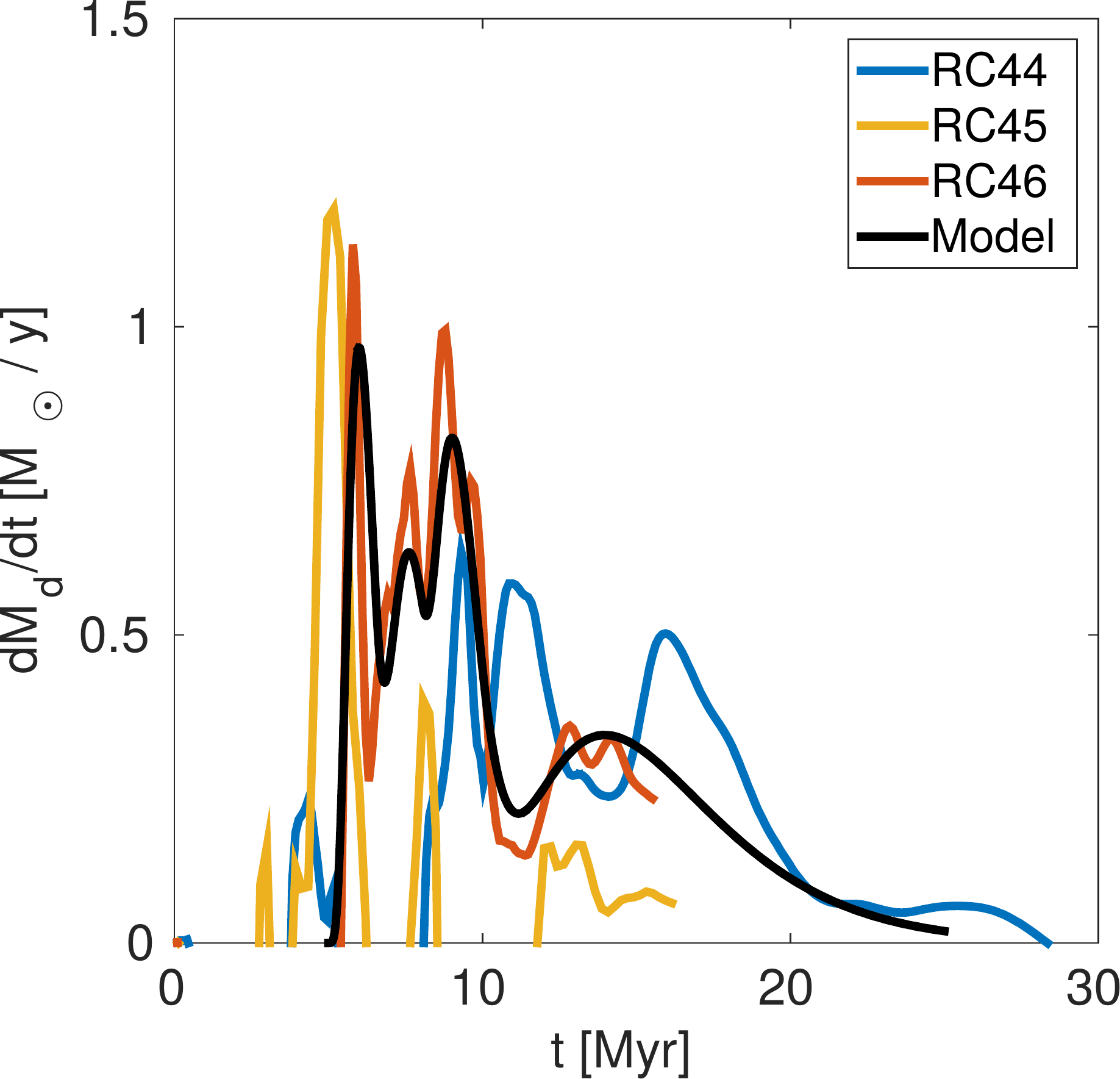}
\end{flushright}
\caption{Mass flux through a cylindrical layer around the central region, radius $2\, kpc h^{-1}$, height $0.2\, kpc h^{-1}$. The coloured curves are for the EC runs (top) and the RC runs (bottom), as indicated by the key. The mass flux is higher by a factor of a few in the RC case. The black curve is a model backflow mass flux inspired by the RC simulations; this is the model we will adopt as mass flow profile in Section \ref{sec:MADdisk}.}
\label{fig:massflux}
\end{figure}

In Figure \ref{fig:massflux} we show the mass flux\footnote{In our notation, a positive flux means inflow} through the same cylindrical region used to estimate $r_c$.

We consider only the net flux through the side surface of the cylinder, as in most cases the flux through the bases is dominated by the jet outflow. This selection may miss some of the bent backflows at the lobe base, thus also in this respect we are just putting lower limits to the mass flux.

{\corr
The flows in Figure \ref{fig:massflux} start with negative values (i.e. there is net outflow, not plotted) due to the predominance of the jet outflow, which still pushes the initial dense environment gas outwards for the first $5-10$~Myr.}

All RC runs present several flux peaks between $5$ and $15$~Myr, reaching about $1\,\mathrm{M_{\odot}/y}$; these peaks originated from the patchy nature of the backflows, but in the RC44 and RC46 cases, {\corr we can see a background flux of about $0.5\,\mathrm{M_{\odot}/y}$}
\corr{until} $15$ or $20$~Myr.  
As noted in C14, similar values could indeed provide substantial central gas accretion, which could contribute to establish a jet self-regulation mechanism;
{\corr this is the subject of}
Section \ref{sec:MADdisk}.

In comparison, {\corr backflows in the EC case} involve masses smaller by a factor of a few (around $0.25\,\mathrm{M_{\odot}/y}$. They also tend to peak at later times ($10$ to $30$~Myr), possibly because of their different morphology which makes the backflow gas traverse a larger distance before approaching the central region.

\section{Parsec-scale backflows and accretion disc kinematics}\label{sec:MADdisk}

In order to estimate the impact of the backflow on the accretion region, in this section we present a model of the backflow-accretion disc interaction, which extends the analysis to 
{\corr scales too small to be reproduced in our}
numerical experiments.

\begin{figure}
\includegraphics[width=\columnwidth]{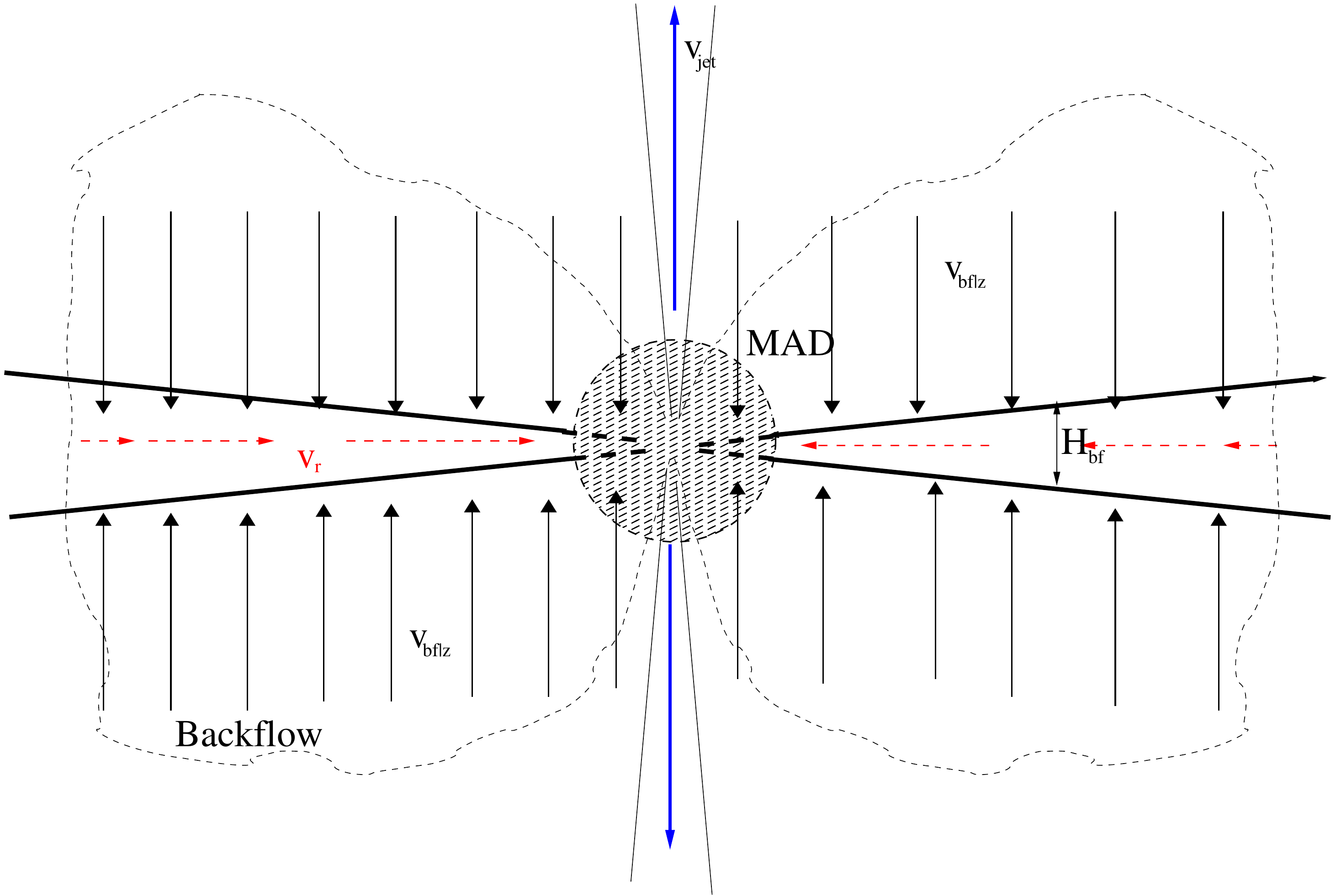}
\caption{Model of the backflow and its action on the central accretion region, the latter being modelled as a \mad. The symmetric backflow reaches the $z=0$ plane and accumulates into a thin disc of height ${\rm H}_{bf}(r) \ll {\rm H}_{\mad}$, where ${\rm H}_{\mad}$ denotes the height of the thick disc. The inflowing accretion flow is confined to the disc (red dotted arrows), and we assume that a mostly azimuthal magnetic field $B_{z}$ (not shown) is present within the \mad region.}
\label{fig:bckfl_model}
\end{figure}

In this model, the central accretion region is assumed to host a \emph{magnetically arrested disc} (hereafter \mad, see Figure~\ref{fig:bckfl_model}).
{\corr The \mad occupies a small region around the centre of a circumnuclear-nuclear disc in the $z=0$ plane, around which the backflows accumulate.}

However, {\corr some further analysis is required to prove that backflows are capable of interacting with the \mad, as the} backflows we observe in the numerical experiments presented above have typically much smaller densities and much larger velocities than in a typical \mad. Our hot and sparse backflows have thermodynamic properties similar to those of \emph{winds}, thus one may ask whether they will effectively accrete onto a disc rather than simply flowing past the $z=0$ plane.

In order to answer this question we will adopt an up-to-date model of the \mad modelled after recent GRMHD simulations, where a slim disc  \citep{1988ApJ...332..646A} {\corr is} threaded by an external magnetic field \citep{1974ApSS..28...45B, 1976ApSS..42..401B,2003PASJ...55L..69N} \citep[see][for a recent review about \mad]{2015ASSL..414...45T}. 
 
{\corr
The \mad is characterized by a \emph{magnetopause}, defined as the region where magnetic and disc thermal pressures are comparable.
In our case the magnetopause can be modelled as a sphere \citep{2003PASJ...55L..69N} of radius $r_{mp} \simeq $ \racc, where \racc $ \simeq 2GM_{BH}/\left(v_{a}^{2}+c_{s}^{2}\right)$, is the black hole's \emph{accretion radius} and $v_{a}$ and $c_{s}$ are the Alfv\'{e}n and sound speed, respectively.
}
The backflow represents an additional flux component within the \mad. \citet{1976ApJ...207..914A} have shown that a wind reaching the magnetopause is likely to be affected by the \emph{exchange instability}, which modifies the morphology of the flow by creating 
knot-like structures aligned along the magnetic field lines. The transverse and longitudinal sizes {\corr of these knots} are given by:
\begin{eqnarray*}
\delta r & \approx & 0.22\sqrt{\frac{GM_{bh}}{R}} t_{ex}, \\
l & \approx & \delta r m_{0}^{1/2},
\end{eqnarray*}
where \corr{$M_{bh}, R,$ and $m_{0}$ are the Black Hole mass, an independent distance variable, and the mode number of the exchange instability{\corr\footnote{For our estimate, we can safely take $ m_0=1$}}, respectively. The characteristic time scale of this instability is {\corr instead} given by}:
\be 
t_{ex}\approx 6.34\times 10^{5} L_{44}^{-1}R_{s}^{3/2}M_{8}^{1/2}\left(15/\ln\left(\Lambda\right)\right)T_{9}^{3/2}\, \, \, {\rm yr }   \label{eq:t_ex}
\ee
\noindent Here $L_{44}$ is the BH luminosity in units of $10^{44}$~erg s$^{-1}$, $R_{s}$ is the \corr{standing shock's} distance\footnote{In the Arons and Lea model the infalling gas is hypersonic and forms a standing shock at a distance $R_{s}$.} in pc, $M_{8}$ is the BH mass (in units of $10^{8}\rm{M}_{\sun}$), $\Lambda$ is the logarithmic Coulomb factor and $T_{9}$ {\corr $\sim [0.01,\, 1]$} is the backflow temperature in units of $10^{9}\,$ K.  
\noindent We can easily check that $t_{ex}\ll t_{ff}$, thus at the magnetopause the exchange instability is effective on the backflow: blobs of typical sizes $\delta r, l$ are created {\corr and} will fall towards the $z=0$ plane, flowing in between the poloidal magnetic field lines. The typical mass {\corr of such blobs} will be: $m_{ex} \approx \rho_{b} \pi l^{2}\delta r$.

Near the central plane the \mad model predicts an almost azimuthal magnetic field, and the backflow blobs, being fully ionized, will become diamagnetic, a feature which will shield them from the further action of the magnetic fields and ease their survival along their path after having crossed the magnetosphere. Moreover, they will also feel a drag force with a characteristic time scale: \citep{1993MNRAS.261..144K,1998ApJ...503..350V}:
\begin{eqnarray} \label{eq:dragtime}
	t_{d} & = & \frac{0.1 v_{A}\rho_{b} \pi l}{B^{2}} \nonumber \\
		& \sim & 3.54\times 10^{4} \mu^{1/2} 
		n_{-2}^{1/2} B^{-1} l \, sec
	\label{eq:tdrag}
\end{eqnarray}
Here \corr{$v_{A}$ and $\mu\simeq 0.62$ are the Alfv\'{e}n velocity and mean molecular weight of a fully ionized hydrogen plasma, respectively, and}: $n_{-2} = n/10^{-2}\, cm^{-3}$ {\corr. $B$ is measured in gauss and $l$ in parsec. In deriving Equation (\ref{eq:dragtime}),} we have assumed that the blobs can be modelled as cylinders of aspect ratio $\delta r/l=0.1$. We thus see that the backflow will be accreted onto the meridional disc on a very short time-scale. As is shown in Fig. 1 of \citet{1998ApJ...503..350V} this drag force is directed in the opposite direction to the blob magnetic field, and  thus it will tend to further drive the blob away from the magnetic field lines.

{\corr We can now proceed to analyse the properties of the backflow-accreting disc.} At variance with a standard thin disc, the accreting gas from the backflow is very hot ($T \gtrsim 10^{8-9}$~K), thus it will be fully ionized and will carry a frozen-in magnetic field. Taking typical values for the backflows from the numerical experiments presented in the previous sections, the internal and kinetic pressures of gas accreting on the disc are:
\begin{eqnarray*}
p_{gas} & \simeq & 1.38\times 10^{-9}\mu n_{-2} T_{9}, \\
p_{kin} \equiv \rho v^{2}  & \simeq & 1.672 \times 10^{-10}\mu  n_{-2} v_{3}^{2}
\label{eq:pres_kin}
\end{eqnarray*}
(in cgs units), where $n_{-2}, v_{3}\,$ and $T_{9}$ are the gas density, velocity and temperature in units of $10^{-2}\, {\rm cm}^{-3}\,$, $10^{3}\, {\rm km/sec}\,$ and $10^{9}\,$ K, respectively. Thus the internal and kinetic pressures are comparable, and the accreted disc will be hot even under the very conservative hypothesis  that no fraction of the kinetic pressure would be dissipated  and converted into internal energy.\\
\noindent The continuity equation for the surface density $\Sigma(r) = \int dz\rho$ in cylindrical coordinates can be written as:
\be
\frac{\partial {\Sigma}}{\partial t} 
+ v_{r}\frac{\partial {\Sigma}}{\partial r}+\frac{\Sigma}{r}\frac{\partial}{\partial r}\left( rv_{r}\right)=\left[ -2\rho_{bf}v_{bf\mid z}\right]_{\mid{z=H/2}}\equiv A(t)
\label{eq:sigma} 
\ee
where the source term in the right-hand  side denotes the mass flux contributed by the backflow. We now assume that the radial inflow velocity is given by:
\be
v_{r} = -\frac{\beta A(t)}{r}
\label{eq:vr}
\ee
This velocity profile is predicted by magnetized Keplerian disc models \citep{1986MNRAS.220..321K}, which also predict an equilibrium surface density profile having the same radial dependence $\Sigma\propto r^{-1}$. Note that this assumption amounts to reducing the time dependence of the radial velocity to that of $A(t)$, the mass flux rate, \corr{implying then}  an instantaneous rearrangement of the accretion flow in the whole disc. We  now assume that an initial low{\corr--}mass, magnetized and Keplerian thin disc exists before backflow infall (i.e. for t$\leq 5\,$ Myr in our experiments), with an initial density profile as given by Kaburaki:
\be
\Sigma_{0}\left(r\right) = \sigma_{0}\frac{r_{0}}{r}.
\label{eq:sigma0}
\ee
Under these conditions, it is easy to show (Antonuccio-Delogu et al, in preparation) that eq.~\ref{eq:sigma} admits an \emph{exact} solution:

\be 
\Sigma(r,t) = \int_{0}^{t}d\tau A\left(\tau\right) + \Sigma_{0}\left(\sqrt[]{{r^{2}}+2\beta\int_{0}^{t}d\tau A\left(\tau\right)}\right)
\label{eq:sigma:soln}
\ee

\begin{table}
\renewcommand{\arraystretch}{1.5} 
\setlength{\tabcolsep}{5pt} 
\begin{center}
	\begin{tabular}{lccccc}
	\hline\hline
i 	&$\dot{\sigma}_{0}$ 			& q	&$\tau_0$	& $\tau_1$	& t$_{s}$\\
		&[$M_\odot/pc^{2}y$]	&  	& [Myr] 	&[Myr]	& [Myr]	\\
\hline		
1		& $0.14$				& $5$			&$0.25$	&$0.2$	&	$5$	\\
2		&$3\times 10^{-4}$				& $3$			&$3.5\times 10^{-2}$	& $0.4$		& $6.5$	\\
3		&$2.5\times 10^{-5}$				&$3$			&$1.5\times 10^{-2}$	&$0.4$		&	$8$ \\
4		&$3.5\times 10^{-5}$				&$2$			&$1.5\times 10^{-2}$	&$2$		&	$10$ \\
\hline		
\end{tabular}
\caption{Parameters of the backflow mass profile model: the linear combination is shown as the template (black continuous line) in Figure \ref{fig:massflux}.}
\smallskip
\label{tab:model}
\end{center}
\end{table}

\noindent The first term on the r.h.s. is always increasing with time: it describes homogeneous physical accretion, independent of position. On the other hand, the second term, for a monotonically decreasing initial density profile such as that   considered here, is  decreasing at any $r$. Physically, this is a consequence of the presence of an inflow (finite $\beta$) which tends to subtract mass, driving it towards the BH. The temporal evolution of the density profile will thus locally depend on the combination of these two opposing 
{\corr terms}.\\
\noindent In Figure~\ref{fig:massflux}, we present the mass variation within a cylindrical surface of radius $r=2\,$ kpc and height 0.1 times the radius. We restrict ourselves only to positive values, i.e. {\corr inflow towards the central BH contributed from} the backflow. {\corr We approximate the total backflow flux as the composition of few accretion episodes}. The black curve in this figure presents a template model obtained by combining four profiles of the form:
\be
A_{i}(t) = \dot{\sigma}_{0}\left(\frac{t-t_{s}}{\tau_{0}}\right)^{q}\exp\left(-\frac{t-t_{s}}{\tau_{1}}\right)\vartheta \left({t_s}\right)
\label{eq:a_t}
\ee
where $\vartheta$ is the Heaviside step function. The parameters of this fit are given in Table~\ref{tab:model}.

To determine the normalization constant $\beta$ we  assume that the accretion radial inflow in the $z=0$ plane is everywhere subsonic: $v_{r} \leq c_{s}$, and that it reaches the sound speed precisely at the \emph{Innermost Stable Circular Orbit}.\\ 
\noindent Using {\corr E}q.~(\ref{eq:vr}) we get: $\beta \leq r c_{s}A^{-1}_{max}$, where: $c_{s}=\left( k T/\mu m_{p}\right)^{1/2}$ is the sound speed and $A_{max}=max_{t\leq t_{max}}\left[ A(t)\right]$, {\corr where we   have set $t_{max} = 20\, Myr$ (typical accretion duration from Figure~\ref{fig:bckfl_model})}. 
Assuming a solar composition plasma ($\mu^{-1}=1.613$) we finally get: $\beta=2.925\times 10^{-3}\left(T_{9}/\mu\right)^{1/2}$ pc$^{4}$ M$_{\odot}^{-1}$, where $T_{9} = T/10^{9}$ K is the only free parameter characterizing the backflow, within this model.\\
\begin{figure}
\includegraphics[width=\columnwidth]{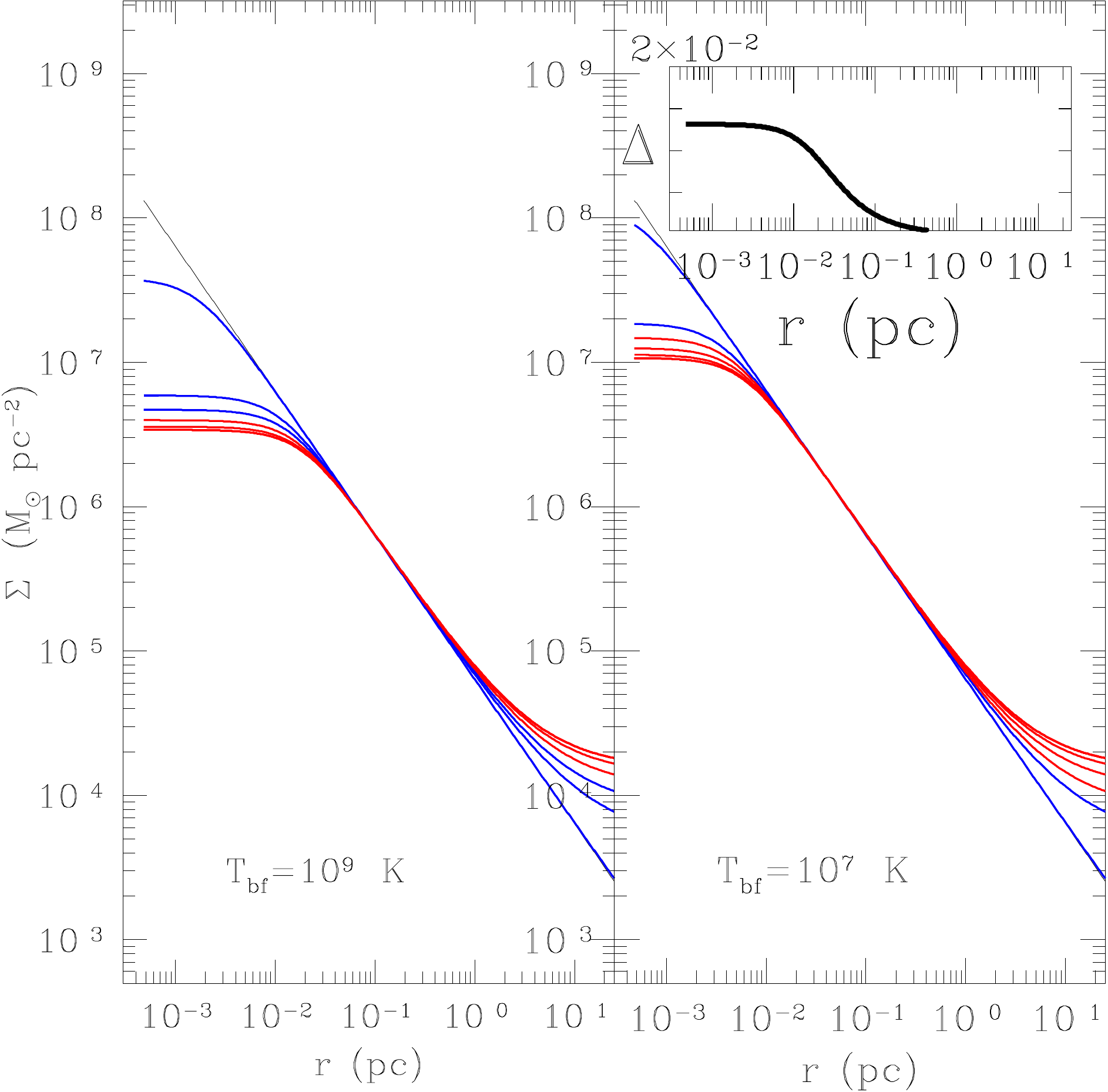}
\caption{Density profile evolution after backflow accretion, for two different backflow temperatures: $\rm{T}_{bf}=10^{9}$ and $10^{7}$ K, respectively left and right. The initial profiles (at $t\leq 5$ Myr) are shown in black, those at early time intervals ($t = 7.5, 10, 12.5$ Myr.) in blue, and the later time profiles ($t = 15, 17.5, 20$ Myr.) in red. The small insertion in the left plot is the global sensitivity $\Delta$ (eq. \ref{eq:sens}). Initial disc with the same M$_{d}$ as in \citep{1973blho.conf..343N} (massive case). We assume a sub-Eddington accretion $\dot{\rm m}_{0} = 0.1$, and a central SMBH mass $\rm{M}_{BH}=10^{8}\, \rm{M}_{\sun}$. }
\label{fig:sigma_ev}
\end{figure}
\noindent In Figure~\ref{fig:sigma_ev} we show the evolution of the density profile for two different values of $T_{9}$. In general, backflow accretion breaks the self-similarity of the density profile
{\corr by creating} 
a flattened tail and a central core. 
This latter feature arises as a result of the competition between accretion from the backflow and an increase 
of the radial inflow towards the BH (eq.~\ref{eq:vr}). At large distances, the flattened density profile is a consequence of the convergence of the integral of $A(t)$ for $t\rightarrow\infty$ to a value which dominates the argument of the second term. This erases any memory of the initial density profile.\\
How robust is the solution given above, and in particular its asymptotic state? We have performed a \emph{sensitivity analysis}  w.r.t. $T_{9}$. In the insert in Figure~\ref{fig:sigma_ev} \corr{(right)} we plot the \emph{total sensitivity} \citep[][eqs. 12 and 17]{1976JCoPh..21..123D} which, for a single parameter, reduces to {\corr the parameter}:
\be
\Delta = \frac{\Delta T_{9}}{\Sigma}\frac{\partial \Sigma}{\partial T_{9}}    \label{eq:sens}
\ee
The total sensitivity gives a measure of the relative variation of $\Sigma$ for a given relative variation of $T_{9}$. As we see from Figure~\ref{fig:sigma_ev}, the most sensible relative variations are limited to the {\corr innermost few $10^{-2}$~pc, i.e. the region occupied by (or closest to) the} central \mad: for larger radii, the final density profile is almost insensitive to variations in the temperature of the backflow, which is directly related to the accretion rate $\dot{m}$. 
\noindent These properties are physically consistent with a further, interesting feature of the exact solution given above (eq.~\ref{eq:sigma}): {\corr the source term $A(t)$ appears in the solution (\ref{eq:sigma:soln}) only under integral, i.e. as}
the \emph{total} accreted mass from the backflow. Thus 
{\corr the final state of the disc} 
($t\rightarrow\infty$) will be \emph{independent} of the time history of backflow.

\section{BH accretion and implications for jet power}\label{sec:self-regulation}
In the magnetically arrested disc (\mad) model there is a tight relationship between
accretion and jet production, the latter being promoted through the Blandford-Znajek mechanism \citep[hereafter BZ][]{1977MNRAS.179..433B}. The details of
this process involve physical mechanisms such as magnetic reconnection
which are not quantitatively understood (see \citealp{2015ASSL..414...45T,2015ASSL..414..149P}  for recent reviews). There is however general  agreement that in the BZ mechanism the jet mechanical power $\rm{P}_{jet}$ scales as $\rm{P}_{jet}\propto\Phi_m^{2}\omega_{d}^{2}$, where $\Phi_m$ is the magnetic flux within the magnetosphere and $\omega_{d}$ is a
function only of the black hole's spin $a$ \citep[see][sect. 3.4]{2015ASSL..414...45T}.

In this section ,we adopt the mass inflow from the circumnuclear disc models described in Section \ref{sec:centralDisk} to estimate the resulting magnetic flux $\Phi_m$ seen by the BH magnetosphere. This calculation comes with a caveat: while the physical scale of interest for a \mad is of order $\simeq100$ Schwarzschild radii, our circumnuclear disc model does not include MHD physics relevant on these scales, as it is instead smoothed down to $r=0$. Thus on intermediate scales 
the physical processes we do not consider here may introduce \emph{time delays} in the accretion (arising from the MHD properties of the disc), or deviate part of the backflow mass into other forms of outflows. We will investigate the effects of some of these features in future, more systematical models of circumnuclear disc/\mad accretion (Antonuccio-Delogu et al. 2016, in prep.).

This first simple, exactly solvable model will give us a first-order estimate about how backflow accretion modifies $\rm{P}_{jet}$. In this section we will simply assume that backflow accretion does not modify the BH mass and spin, and only affects the magnetic flux within the BH magnetosphere.
This is a reasonable assumption, as accretion of hot, sparse gas should have little 
effect on the growth of BH properties.

The magnetosphere of the BH is defined as the region where magnetic pressure
and gravitational attraction are comparable:
\be
\frac{B_{m}^{2}}{8\pi} = \frac{GM_{BH}\Sigma_{m}}{r_{m}^{2}} \label{pj:1}
\ee
where $\Sigma_{m}\equiv\Sigma(r_{m})$.  The size of the magnetic field threading the BH-disc system is not uniquely determined by the physics of accretion: however we know that the \mad region ($r\leq r_{m}$) will also be threaded by the external magnetic field which threads the disc, and it will have a density different from that of the disc itself. We assume that at the magnetopause discontinuity, the Alfv\'{e}n velocity $v_{A} = \left(B^{2}/4\pi\rho\right)^{1/2}$ will be continuous, which is equivalent to assuming that the magnetosphere will be in an equilibrium thermal state, and will not be heated by Alfv\'{e}n waves diffracted at $r_{m}$. Thus we have: $B\propto\rho^{1/2}$, and taking into account the  disc density profile given before ($\rho\propto r^{-1}$), we eventually obtain:
\be
B(r) = B_{m}\left(\frac{r_{m}}{r}\right)^{1/2}  \label{pj:2}
\ee
Thus, the magnetic flux within the magnetosphere will be given by:
\[
\Phi_{m} = \int_{r_{g}}^{r_{m}}dr 2\pi rB(r) \backsimeq \frac{4\pi}{3}B_{m}r_{m}^{2}     
\]
($r_{m}\gg r_{g}$) and, substituting $B_{m}$ from eq~\ref{pj:1} above, we eventually get:
\be
 \Phi_{m} = \left(\frac{128\pi^{2}G}{9}\right)^{1/2}M_{BH}^{1/2}r_{m}\Sigma_{m}^{1/2}  \label{pj:3}
\ee
The accretion rate onto the BH is defined as: $\dot{M} = 2\pi r_{m}\epsilon v_{ff} = 2\pi\epsilon\left( GM_{BH}r_{m}\right)^{1/2}$, where we have used the definition of free-fall velocity ($v_{ff}=\left(GM_{BH}/r_{m}\right)^{1/2}$). 
\begin{figure}
\includegraphics[width=\columnwidth]{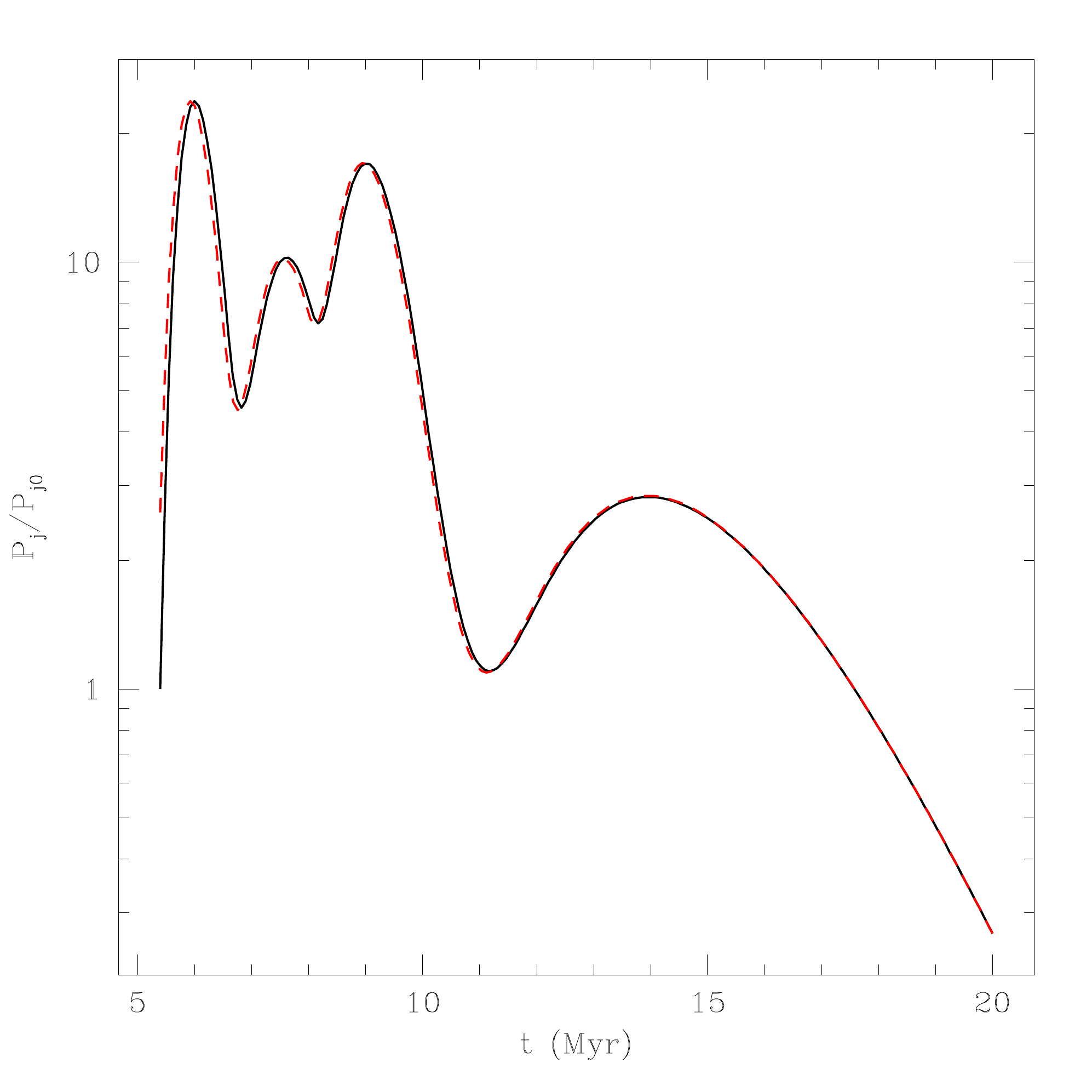}
\caption{Temporal evolution of the jet's relative power $P_{j}/P_{j0}$, where the denominator is the average power when the backflow is not taken into account. The continuous line corresponds to the same initial profile shown in Figure \ref{fig:sigma_ev} for $T = 10^{9}\, {\rm K}$; the dashed red line is for $T = 10^{7}\, {\rm K}$. The almost similar evolution shows that the accretion history is mostly driven by the \emph{dynamical} evolution of the backflow.}
\label{fig:pjet_ev}
\end{figure}
Here $\epsilon \backsimeq  0.1 \leq 1$ is a fudge factor which quantifies the average efficiency of accretion. Defining as usual the accretion rate in units of the Eddington accretion rate: $\dot{m}\equiv\dot{M}/\left(\eta M_{BH}\right)$, we finally arrive at an expression for $r_{m}$ as a function of the accretion rate:
\be
r_{m} = \left(\frac{\eta}{2\pi}\right)^{2}G^{-1}\frac{M_{BH}\dot{m}^{2}}{\epsilon^{2}\Sigma_{m}^{2}}    \label{pj:4}
\ee
Inserting this expression into eq.~\ref{pj:3} we get:
\be
\Phi_{m} = \kappa G^{-1/2}M_{BH}^{3/2}\dot{m}^{2}\epsilon^{-2}\Sigma_{m}^{-3/2} \label{pj:5}
\ee
where $\kappa$ is a purely numerical constant. To proceed further, we note that the backflow will modify the accretion rate at $r_{m}$ as given by eq.~\ref{eq:vr} above, thus we will have:
\[
\dot{m}=\frac{\dot{M}}{\dot{M}_{Edd}}=\frac{4\pi r_{m}^{2}\rho(r_{m})v_{r}}{\eta M_{BH}}    
\]
and, using: $\rho(r_{m})=\Sigma_{m}/\left( r_{m}/2\right)$ and the expression for $v_{r}$ we eventually find that the reduced accretion rate during the backflow scales as: $\dot{m}\propto M_{BH}^{-1}\Sigma_{m}A(t)$. After having substituted in eq.~\ref{pj:5}, we eventually obtain:
\be
\Phi \propto M_{BH}^{-1/2}\epsilon^{-2}\Sigma_{m}^{1/2}A^{2}(t)    \label{pj:6}
\ee
We plot in Figure~\ref{fig:pjet_ev} the temporal evolution of the jet's mechanical power, in units of the (constant) power without backflow. The similarity with the evolution of the azimuthal flux in Figure~\ref{fig:massflux} is a consequence of the hypothesis that the mass flux variation (the source term in eq.~\ref{eq:sigma})  generates variations of $\Sigma$ without any delay and/or modifications: the latter could arise for instance by finite conductivity effects which can threaten the ideal coupling between streamlines and magnetic fields, or by the enhancement of magnetic reconnection \citep[see e.g.][]{2015ASSL..414..149P} when the backflow plasma carries a magnetic field. Thus, the results we have presented in this section represent only the simplest possible scenario, and we will consider more realistic ones in future papers (Antonuccio-Delogu et al. 2016, in prep.).

\section{Conclusion}

Backflows are a large-scale generic feature of jet propagation within the ISM of their host galaxies. The high-resolution numerical 3D experiments we have presented here confirm our previous findings concerning their physical origin \citep{2010MNRAS.405.1303A}, and in particular concerning their \emph{thermodynamic} origin: it is the discontinuity in enthalpy near the hotspot, together with the finite curvature radius of this discontinuity, that generates a finite vorticity in the initially laminar jet flow, as predicted by \emph{Crocco's theorem} \citep{1937ZaMM...17....1C}.\\
There is also  mounting observational evidence for backflows from analysis of FR-I radiogalaxies \citep{laing2012backflows}, and from more recent studies concerning Cen-A:
{\corrx
\cite{neumayer2007cenA} observed inflow in the kinematics of highly ionized gas in the nucleus of the source. They associate this feature with backflow of gas that was accelerated by the jet of Cen-A (hence the high ionization rate), confirming backflow presence down to parsec scales, as later confirmed by further studies
\citep{2015A&A...575L...3H, 2013ApJ...766...36B}. 
}
Until now, however, the most complete evidence for backflows has emerged from previous numerical simulations similar to those presented here  \citep{2014MNRAS.437.3405L, 2007MNRAS.382..526P}, and also from simulations including a very inhomogeneous medium \citep{2012ApJ...757..136W, 2011ApJ...728...29W}: in the latter, the backflow mostly originates from jet gas flows that have run into cold clouds and thus have an intermittent character. The high resolution, adaptive numerical 3D experiments we have presented here demonstrate instead that backflows are \emph{spatially coherent} and can have a significant \emph{temporal extension}, \corr{lasting for few tens of million years} : both are \emph{structural} features of jet propagation within their host galaxies.\\
The backflows will be fed as long as there are the physical conditions to generate them and support a finite curvature enthalpy discontinuity near the hotspot, either in the form of a shock or of a contact discontinuity. These features depend only slightly on \pjet, as in most cases we observe
a highly intermittent ($\dot{M}\simeq 10^{-1}-1\, \rm{M}_{\sun} \rm{yr}^{-1}$) inflow towards the central BH region lasting about 15-20 Myrs. The morphology of the cavities may nonetheless change these  numbers by a factor  of order two (as seen from the differences between our EC and RC series, e.g. comparing Figures~\ref{fig:diskEC} and ~\ref{fig:diskRC}).
Due to its axial symmetry, the backflow has a low azimuthal angular momentum w.r.t. the central accretion region. The combination of this circumstance and of the thermodynamics of its propagation conspire to drive a fraction of the backflow  towards the very central regions.\\
We have explored the impact of backflows on the central unresolved accretion region near the supermassive BH, and in particular on the jet power, assuming backflow can be modelled as a hot gas accretion onto a \mad, \corr{this latter being} a paradigm supported by many GRMHD simulations \citep{2011ApJ...728L..17P,2010MNRAS.402..497G}. The analysis we presented in Sect. 5 is not free of some assumptions. For instance, we have disregarded the dissipative processes in the innermost regions of the accretion disc, which will heavily affect the magnetic field strength and topology and dominate over the action of the time-dependent accretion flow. In our model the backflow will accrete into a thin, hot, high-$\beta$ disc, and we have shown that the jet power \pjet can be boosted by a factor of 10 or 20 for as long as 5 or 10 Myr. The temporal evolution of the disc density and jet power in this model is dependent on the profile and normalization of the initial disc: however, the final density profile will be independent of the details of the temporal evolution of backflow accretion. As is clear from Figure~\ref{fig:massflux}, all of  the mass accretion episodes start about 3-5 Myr. after jet launching and decay after 20-30 Myr, independently of \pjet. However, as is clear from a visual inspection of Figure~\ref{fig:sigma_ev}, despite large individual variations between different simulations, our "template" model reproduces the generic features of accretion episodes.\\
From our analysis, it is evident that the backflow is the result of the interaction between the jet and the local host galaxy's interstellar environment, and its contribution to the \mad 
demonstrates that a connection between \emph{galaxy-scale feedback}
and \emph{central accretion} inevitably develops on time-scales of the order
of $\sim10^{6}$ years, or about $1/10$\, of the AGN duty cycle. This \emph{backflow accretion } time -scale however only refers to the typical time for the backflow to feed the \mad. The backflow  phenomenon points to a deep connection between AGN feedback and SMBH accretion, as previously hinted at \citep{2008NewAR..51..733N}
\footnote{Although magnetic phenomena may introduce further time delays, as briefly mentioned in Sec.~\ref{sec:self-regulation}}.\\
\noindent Finally, we would like to point out  that the backflow is a significant \emph{global dynamical feature} of an AGN that is capable of "bridging" the very large (kpc) scales, where jets propagate, with the accretion (subparsec) scales. The "feedback" from the large scales is capable of modulating \emph{global} properties of the jet, such as  the mechanical power P$_{jet}$, which in turn can affect the thermodynamic properties near the hotspot, from where the backflow originates. This cycle creates a \emph{self-regulation} mechanism which determines the duty cycle and other properties of the AGN, as we will show in a separate paper (Antonuccio-Delogu et al., in preparation).

\section*{Acknowledgements}
We would like to acknowledge the anonymous referee for the careful
reading and useful comments which improved the quality of the paper.\\
The work of S.C. has been supported by the ERC Project No. 267117 (DARK) hosted by Universit\'{e} Pierre et Marie Curie (UPMC) - Paris 6 and the ERC Project No. 614199 (BLACK) at Centre National De La Recherche Scientifique (CNRS). SC thanks Marta Volonteri for the profitable discussion and the precious advice.\\
\noindent The work of V.A.-D. \dots has partially been supported by the joint CNRS-INAF Project PICS 2013-2016 "\emph{Modelling and Simulation of mechanical AGN Feedback}". V.A.-D. gratefully acknowledges the hospitality of IAP, Paris, during the completion of this work, particularly very useful conversations with G. Mamon and M. Volonteri.
The work of 
JS was supported by ERC Project No. 267117 (DARK)
hosted by Universit\'{e} Pierre et Marie Curie (UPMC) - Paris
6, PI J. Silk.
\noindent The software used in this work was in part developed by the DOE NNSA-ASC OASCR Flash Centre at the University of Chicago.


\bibliographystyle{mn2e}
\addcontentsline{toc}{section}{\refname}


\appendix

\section{Numerical algorithms and code} \label{appendix}
The AMR code we have chosen to perform the numerical experiments described in this  work (FLASH v. 4.2)  solves the Euler equations system:
 \be 
\frac{\partial\rho}{\partial t} + \nabla\cdot\left( \rho\mathbf{v}\right) = 0     \label{eq:eu1}
\ee 
\be 
 \frac{\partial\rho\mathbf{v}}{\partial t} + \nabla\cdot\left( \rho\mathbf{v}\mathbf{v}\right) + \nabla P = \nabla\left(\rho \mathbf{g}\right)  \label{eq:eu2}
 \ee 
 \be
  \frac{\partial\rho E}{\partial t} + \nabla\cdot\left[\left(\rho E + P\right)\mathbf{v}\right] = \rho \mathbf{v}\cdot\mathbf{g} -\Lambda\left(\rho, T\right)\label{eq:eu3}
 \ee
 where: $E=\epsilon + v^{2}/2$ and $\epsilon$ is the specific internal energy, $P$ is the (thermal) pressure and $\Lambda$ the cooling function. 
 
Among the different numerical algorithms made available in FLASH we have adopted the shock-capturing PPM scheme \citep{1984JCoPh..54..115W} which is particularly suited to model the shocks formed during the interaction of jets with the cocoon. 
 
We use FLASH's default \emph{ideal gamma} equation of state for an ideal gas, $P=(\gamma-1)\rho\epsilon$, where $\gamma$ is the specific heat ratio ($5/3$ in our case).
As mentioned in the text, we include a radiative cooling function extended to very high $T \geq 5\times 10^{9}$ K extended to very high temperatures to account for radiative losses due to $e^{+}-e^{-}$ annihilations \citep[see the Appendix of][]{2008MNRAS.389.1750A}.

We adopt FLASH's \emph{Multigrid Poisson Gravity Solver}, with a custom modification for adding a static dark matter gravitational potential for the host NFW halo.

FLASH adopts the PARAMESH block-structured AMR decomposition of the spatial  computational domain \citep{MACNEICE2000330}. In PARAMESH the initial grid is recursively refined dichotomically up to a maximum refinement level $l_{max}$. 
Moreover, each block is further divided into $n_c$ computational cells along each spatial direction. Thus, the minimum resolved block of cells has a linear size: $\Delta r=L/2^{l_{max}} n_c$. In all our runs we have:  $\Delta r=\frac{640\ \mathrm{kpc}}{2^{10} \times 8} = 78.125$~pc.

Finally, we allow refinements on the basis of a density and temperature gradient criterion, adopting FLASH default refinement strategy based on Loehner's error estimator \citep[][see FLASH user manual]{lohner1987adaptive} set to $0.8$ for refinement and $0.6$ for de-refinement. 
\newpage



\label{lastpage} 
\end{document}